%% file: ris_nc_beamtrain.tex
\documentclass[journal,comsoc]{IEEEtran}

\usepackage[T1]{fontenc}

\ifCLASSINFOpdf
  \usepackage[pdftex]{graphicx}
  \graphicspath{{./images/}}
\else
\fi

\usepackage{hyperref}

\usepackage{mathtools}
\usepackage{mathrsfs}
\usepackage{amsmath}
\usepackage{bm}
\usepackage[cmintegrals]{newtxmath}

\ifCLASSOPTIONcompsoc
  \usepackage[caption=false,font=normalsize,labelfont=sf,textfont=sf]{subfig}
\else
  \usepackage[caption=false,font=footnotesize]{subfig}
\fi
\usepackage{cite}
\usepackage{tabularx}
\usepackage{xcolor}
\usepackage{multirow}

\input{NewCommands.tex}

\begin{document}

\title{Differential Data-Aided Beam Training for RIS-Empowered Multi-Antenna Communications}
	
\author{Kun Chen-Hu,~\IEEEmembership{Member,~IEEE,} George C. Alexandropoulos,~\IEEEmembership{Senior~Member,~IEEE,} \\and Ana García Armada,~\IEEEmembership{Senior~Member,~IEEE}
	
\thanks{K. Chen-Hu and A. García Armada are with the Department of Signal Theory and Communications, Universidad Carlos III de Madrid, 28911 Leganés, Spain (e-mails: \{kchen, agarcia\}@tsc.uc3m.es).}

\thanks{G. C. Alexandropoulos is with the Department of Informatics and Telecommunications, National and Kapodistrian University of Athens, 15784 Athens, Greece (e-mail: alexandg@di.uoa.gr).}

\thanks{This work has been funded by the Spanish National project IRENE-EARTH (PID2020-115323RB-C33 / AEI / 10.13039/501100011033) and the European EU H2020 RISE-6G project under grant number 101017011.}
}


\maketitle

\begin{abstract}
The Reconfigurable Intelligent Surface (RIS) constitutes one of the prominent technologies for the next generation of wireless communications. It is envisioned to enhance the signal coverage in cases when the direct link of the communication is weak. Recently, beam training based on codebook selection is proposed to obtain the optimized phase configuration of the RIS. After that, the data is transmitted and received by using the classical coherent demodulation scheme (CDS). This training approach is able to avoid the large overhead required by the channel sounding process, and it also circumvents complex optimization problems. However, the beam training still requires the transmission of some reference signals to test the different phase configurations of the codebook, and the best codeword is chosen according to the measurement of the received energy of the reference signals. Then, the overhead due to the transmission of reference signals reduces the spectral efficiency. In this paper, a zero overhead beam training for RIS is proposed, relying on data transmission and reception based on non-CDS (NCDS). At the BS, the received differential data can also be used for the determination of the best beam for the RIS. Therefore, the efficiency of the system is significantly enhanced since reference signals are fully avoided. After choosing the best codebook, NCDS is still more suitable to transmit information for high mobility scenarios as compared to the classical CDS. Analytical expressions for the Signal-to-Interference and Noise Ratio (SINR) for the non-coherent RIS-empowered system are presented. Moreover, a detailed comparison between the NCDS and CDS in terms of efficiency and complexity is also given. The extensive computer simulation results verify the accuracy of the presented analysis and showcase that the proposed system outperforms the existing solutions.

\end{abstract}


%
\IEEEpeerreviewmaketitle

\section{Introduction}
\label{sec:introduction}

\IEEEPARstart{T}{he} emerging technology of Reconfigurable Intelligent Surface (RIS) \cite{Ris01,liaskos2018new,di2019smart,Ris03,Ris02,george1,george2} is expected to play a significant role in the evolution of mobile communication systems, from the current 5-th Generation (5G) \cite{nr-211} towards the 6-th Generation (6G) \cite{Samsung}. The high frequency bands will be extensively exploited for mobile communications \cite{nr-901}, such as $3.5$ GHz and millimetre waves (mm-Wave), in order to take advantage of the huge available bandwidth and provide a fully enhanced Mobile Broadband (eMBB) experience. The drawback is that the coverage in these bands will suffer from a strong attenuation loss. RIS-empowered links is an appealing solution to both improving and extending the signal transmitted by either the Base Station (BS) or User Equipment (UE), without excessively increasing the overall cost of the wireless network.

RISs are lightweight and hardware-efficient artificial planar structures of nearly passive reflective elements \cite{Ris02} that enable desired dynamic transformations on the signal propagation environment in wireless communications~\cite{di2019smart}. They can support a wide variety of electromagnetic functionalities \cite{WavePropTCCN,george3}, ranging from perfect and controllable absorption, beam and wavefront shaping to polarization control, broadband pulse delay, radio-coverage extension, and harmonic generation. The RIS technology is envisioned to coat objects in the wireless environment~\cite{di2019smart} (e.g., building facades and room walls), and can operate either as a reconfigurable beyond Snell's law reflector \cite{Ris01}, or as an analog receiver~\cite{hardware2020icassp,HRIS_Nature} or lens~\cite{RISLens2021} when equipped with a single Radio-Frequency (RF) chain, or as a transceiver with multiple relevant RF chains~\cite{DMA2020}.  

The exploitation of RISs requires to obtain a proper phase configuration of its reconfigurable elements, capable of creating an alternative high-gain reflective channel between BS and UE \cite{9053695,8879620,yuanxiaojun_ce,RisChanEst01,RisChanEst02,RisChanEst04,RisChanEst05,george4,newr3_stat}. For that purpose, a significant amount of reference signals is transmitted in the uplink to obtain the so-called cascaded channel estimation, which encompasses the joint effect of the signal propagation over the BS-RIS and RIS-UE links \cite{9053695,8879620,yuanxiaojun_ce}. Note that this overhead not only becomes prohibitive as the numbers of UEs, RIS elements, and configurations increase \cite{Ris04}, but it is absolutely useless when the cascaded channel has a strong attenuation due to the multiplicative fading effects \cite{multifad}. Once the cascaded channel estimation is available at the BS, the best pair of precoder/combiner and phase configuration are computed and communicated to the RIS via a side control link. This processing task is not straightforward due to the fact that a non-convex design optimization needs to be solved, increasing the operational complexity. Several recent studies have focused on accelerating this optimization using alternative methods at the expense of sacrificing the performance with sub-optimal optimizations \cite{Ris01,Ris03,huang2018energy,RisChanEst01,RisChanEst02,RisChanEst03}. After the RIS is adequately configured, the traditional Coherent Demodulation Scheme (CDS) is used for the data transmission. It is generally assumed that the channel coherence time is long enough to encompass the estimation, optimization, and data transmission, a condition that may be difficult to satisfy, especially in mobile communication systems.

Recently, different alternatives have been proposed in order to alleviate the overhead incurred by the cascaded channel estimation and reduce the delay induced by the optimization, including beam training \cite{ris_beam_geo,ris_beam_rui,george5,newr1,newr2} and the exploitation of the diversity gain of the RIS \cite{ris_div_ran, ris_div_nc}. The former approaches \cite{ris_beam_geo,ris_beam_rui,george5,newr1,newr2} use codebooks based on a reduced number of phase configurations, capable of producing several directive beams pointing towards some specific area(s) of interest. Similarly to the beam training proposed for the multi-antenna BS at mm-Wave \cite{beam_chenhao,beam_george}, the UE transmits some reference signals and the BS measures the received signal strength when each codeword is applied. The chosen codeword is that one associated with the highest received energy. This solution is not only able to reduce the required time for channel estimation and optimization, but it also enables this technology to be used at high path-loss scenarios. However, it still requires a beam training period to test all the phase configurations of the codebook. Then, in order to reduce further the number of phase configurations, \cite{ris_beam_rui} proposed the use of multiple beams at the expense of reducing their directivity. However, the exploitation of beam alignment at mm-Wave will arise some security issues related to jamming attacks \cite{newbaja1,newbaja2}. A jammer may transmit high-power signals to intentionally induce a beam failure event, and the UEs are not capable to access the network or switch to a new better beam. On the other hand, the strategies in \cite{ris_div_ran, ris_div_nc} assumed that the phase configurations at the RIS are randomly chosen over time, and hence, the received signal will be enhanced due to the time and/or spatial diversity. However, \cite{ris_div_ran} only provides some theoretical bounds in terms of outage probability and achievable rate, assuming that any modulation and coding scheme can be applied. In turn, \cite{ris_div_nc} proposed the use of Non-CDS (NCDS) \cite{Ana2015,Kun2019,Kun2020,KunJFET} in order to increase the rate, by avoiding the transmission of reference signals and only exploiting the spatial diversity produced by both the BS and RIS. Although the average performance is scaled by the spatial diversity produced by both the antennas of the BS and the passive elements of the RIS, the received signal may suffer from strong fading for certain periods, since some random phase configurations may not be favorable for the UE of interest.

To the best knowledge of the authors, a zero overhead beam training based NCDS for RIS-aided communications (NCDS-RIS) has never been proposed. It would provide the advantages of beam training \cite{ris_beam_geo,ris_beam_rui,george5,newr1,newr2} and NCDS \cite{Ana2015,Kun2019,Kun2020,KunJFET}. During the beam training process, the effective BS-UE channel is strongly time-varying as a consequence of testing different phase configurations of the RIS profile codebook. However, data transmission can be efficiently performed during those training periods by using NCDS \cite{Ana2015,Kun2019,Kun2020,KunJFET}, which allows to both measure the received energy for each chosen codeword and demodulate the received signals without the knowledge of the channel estimation. Unlike CDS, it is very robust to the time variability. Hence, the best codeword can be selected without sacrificing the data-rate. Moreover, after the alternative link between BS and UE via RIS has been established, the NCDS is still a preferable choice in some circumstances as compared to the traditional CDS, as will be shown. For high mobility scenarios where the channel coherence time may be small, the NCDS-RIS is able to fully avoid the estimation of the resulting channel between BS-UE, and hence, the data-rate is increased. Also in this transmission stage, computing the precoder/combiner at the BS is not required, reducing the complexity, delay and energy consumption of the system. Additionally, our proposed scheme are compatible with the different anti-jamming techniques \cite{newbaja1,newbaja2}, such as pseudo-random multi-beam pattern, randomized probing, etc.

Motivated by the above described facts, in this paper, we propose the novel NCDS-RIS, which consists of the combination of RIS beam training and NCDS in order to fully avoid the overhead, targeting moderate to high mobility scenarios for 5G Advanced and 6G systems. The main contributions of the paper are summarized as follows:
\begin{itemize}
	\item An RIS-empowered Single-Input Multiple-Output (SIMO) Orthogonal Frequency-Division Multiplexing (OFDM) system with differential-data-aided beam training is presented, denoted as NCDS-RIS. This combination requires neither a lengthy channel estimation process nor solving a non-convex optimization problem for the RIS configuration. In order to maximize the throughput, the way to encode the non-coherent data based on differential Phase-Shift Keying (PSK) modulation in the time and frequency resources of the OFDM signal is proposed for the different stages of the communication. During the beam training stage, differential modulation is proposed to be exclusively performed in the frequency domain. This is the best solution because the channel will suffer from strong variations in consecutive OFDM symbols due to the beam training process. At this stage, the energy of the received data signal is measured to choose the best phase configuration for the RIS. After the training process is done and the best codeword is chosen to configure the RIS, the differential PSK modulation is proposed to be performed in both frequency and time domains. Consequently, the proposed approach significantly enhances the data-rate of the system for both stages by reducing the required reference signals.
	
	\item The Signal-to-Interference-plus-Noise Ratio (SINR) of the proposed RIS-empowered NCDS system, determining the power of the useful signal over the self-interference and thermal noise, is analytically characterized over a realistic geometric wideband channel model \cite{chan_meas1,chan_meas2,chan_meas3}.
	
	\item The throughput and complexity of the proposed NCDS-RIS system are analyzed and compared to the existing solutions based on reference signals and CDS transmission. The throughput is computed taking into account the amount of overhead required and the bit error rate (BER) of each approach, while the number of complex product operations accounts for the complexity for all the chosen techniques.
	
	\item The simulation results verify the accuracy of the presented SINR analysis and highlight the superiority of the proposed NCDS-RIS system over the state of the art. The BER and throughput are also numerically assessed using the 5G numerology for scenarios with different degrees of mobility. The performance comparison shows the benefits of the proposed NCDS-RIS, relative to the existing CDS-RIS.
\end{itemize}

The remainder of the paper is organized as follows. Section~\ref{sec:system_model} introduces the system model and the geometric wideband channel model. Section \ref{sec:beamtraining} outlines the channel gain of the RIS-aided communication and the codebook design. Section \ref{sec:nc-diff} details the implementation of the proposed differential PSK scheme. Section \ref{sec:sinr} presents the analytical expressions for the SINR. Section~\ref{sec:efficiency} includes the comparison analysis among the proposed NCDS-RIS and the existing solutions based on CDS in terms of throughput and complexity. Section~ \ref{sec:num_res} discusses the performance assessment results. Finally, Section~\ref{sec:conclusion} concludes the paper.

\textbf{Notation:} Matrices, vectors, and scalar quantities are denoted by boldface uppercase, boldface lowercase, and normal letters, respectively. $\left[\mathbf{A}\right]_{mn}$ denotes the element in the $m$-th row and $n$-th column of $\mathbf{A}$.
$\jmath$ is the imaginary unit, $\eabsn{\cdotp}{}$ represents the absolute value and $\measuredangle \ecs{\cdotp}$ corresponds to the phase component of a complex number.
$\enormn{\cdotp}{}{F}$ is the Frobenius norm.
$\ereal{\cdotp}$ and $\eimag{\cdotp}$ represent the real and imaginary parts, respectively.
$\otimes$ and $\odot$ denote the Kronecker and Hadamard products of two matrices, respectively.
$\eexp{\cdotp}$ represents the expected value of a random variable, $\evar{\cdotp}$ denotes the variance, and $\mathcal{CN}(0,\sigma^2)$ represents the circularly-symmetric and zero-mean complex normal distribution with variance $\sigma^2$. $\eexpod{\nu}$ accounts for the Exponential distribution with rate parameter $\nu$.

\eAddTwoColFig{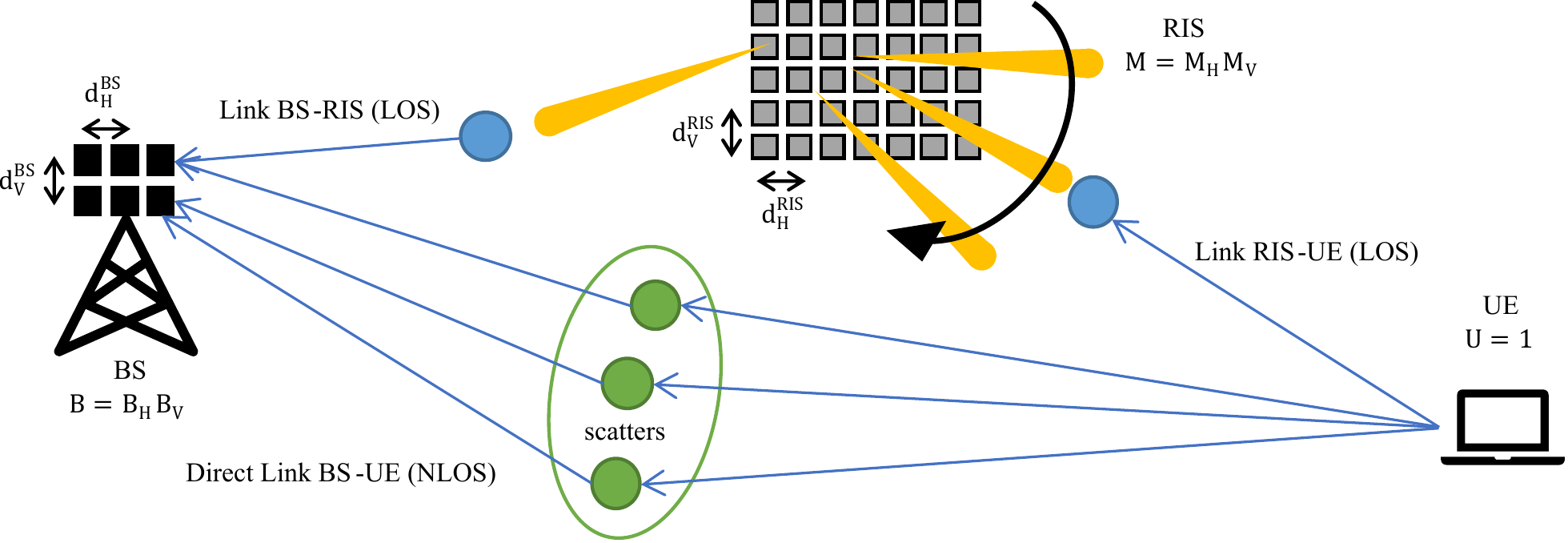}{0.8}{fig:system}{The RIS-empowered wireless communication link comprising a multi-antenna BS, a multi-element passive panel, and a single-antenna mobile UE.}

\section{System and Channel Models}
\label{sec:system_model}

The considered mobile communications scenario comprises a BS, an RIS, and a single-antenna UE (see Fig. \ref{fig:system}). The BS is equipped with a uniform rectangular array (URA) consisting of $B=B_{H}B_{V}$ antenna elements, where $B_{H}$ and $B_{V}$ denote the number of elements in the horizontal and vertical axes, respectively, and the distance between any two contiguous elements in their respective axes is given by $d_{H}^{\text{BS}}$ and $d_{V}^{\text{BS}}$. Analogously to the BS, the RIS is built by $M=M_{H}M_{V}$ fully passive reflecting elements, whose respective distances between elements are given by $d_{H}^{\text{RIS}}$ and $d_{V}^{\text{RIS}}$. We finally assume that the RIS is attached to a dedicated controller for managing its configuration, which is synchronized with the BS.

Focusing on the uplink case, the UE transmits the data to the BS using both the direct BS-UE link and the reflected link via the BS-RIS and RIS-UE communication links. It is understood that other UEs may be multiplexed in different orthogonal (e.g. time or frequency) resources. 

\subsection{Channel Model}
\label{subsec:geo}
All the channel links (BS-UE, BS-RIS, and RIS-UE) are modeled by a geometric wideband channel model \cite{chan_meas1,chan_meas2,chan_meas3}, made up of the superposition of several separate clusters, each of them with a different value of delay, gain, and angles of arrival and departure. The delays and geometrical positions of each cluster/ray are typically characterized by the Delay and Angular Spreads (DS and AS), respectively. The spatial-temporal information of these clusters and the array steering vectors are included to model the spatial correlation due to the array responses. Therefore, this channel model is able to account for the spatial correlation considering both the given antenna array response of the BS and RIS, as well as the geometrical positions of all clusters. 

Moreover, it is assumed that the BS-RIS channel impulse response is time invariant and characterized by Rician fading, since both BS and RIS are fixed elements (e.g., placed at the top of a wall). In turn, the BS-UE and RIS-UE links are considered to remain quasi-static within the channel coherence time ($T_{c}$), due to the mobility of the UE. The direct BS-UE link is modeled as Rayleigh fading, assuming that there are obstacles preventing a line-of-sight (LoS) component. However, the RIS-UE link is modeled as Rician, since it is considered that the reflected link via the RIS is able to establish an alternative LoS link, capable of avoiding the mentioned obstacles present at the direct link. In the considered case where there is not any phase configuration capable of providing better alternative link, the communication is still carried through the direct BS-UE link.

Hence, the channel links at the $k$-th subcarrier (with $k=1,2,\ldots,K$) can be described as
\begin{equation}\label{eqn:ref_bs_ue_all}
	\mathbf{h}_{d,k}  \triangleq \sqrt{L_{d}} \tilde{\mathbf{h}}_{d,k} \ecsizeo{B},
\end{equation}
\begin{equation}\label{eqn:ref_bs_ris_all}
	\mathbf{G}_{e,k}  \triangleq \sqrt{\frac{L_{e}}{\delta_{e}+1}} \left( \sqrt{\delta_{e}}\overline{\mathbf{G}}_{e,k} + \tilde{\mathbf{G}}_{e,k}\right) \ecsize{B}{M}, 
\end{equation}
\begin{equation}\label{eqn:ref_ris_ue_all}
	\mathbf{g}_{u,k}  \triangleq \sqrt{\frac{L_{u}}{\delta_{u}+1}} \left( \sqrt{\delta_{u}}\overline{\mathbf{g}}_{u,k} + \tilde{\mathbf{g}}_{u,k}\right) \ecsizeo{M},
\end{equation}
where $L_{d}$, $L_{e}$, and $L_{u}$ are the large-scale gain for the BS-UE, BS-RIS, and RIS-UE links, respectively, $\delta_{e}$ and $\delta_{u}$ correspond to the Rician factor of the BS-RIS and RIS-UE links, respectively. Note that the overall channel gain is dominated by the large-scale gains, since they are significantly lower than the Rician factors ($L_{e} << \delta_{e}$ and $L_{u} << \delta_{u}$). In addition, $\tilde{\mathbf{h}}_{d,k}$, $\tilde{\mathbf{G}}_{e,k}$, and $\tilde{\mathbf{g}}_{u,k}$ account for the non-LoS (NLoS) components of the channels, while $\overline{\mathbf{G}}_{e,k}$ and $\mathbf{g}_{u,k}$ are the LoS ones.

The channel impulse response of the direct BS-UE link at the $k$-th subcarrier can be described as
\begin{equation}\label{eqn:direct_bs_ue}
	\tilde{\mathbf{h}}_{d,k} \triangleq \sum_{c=1}^{C_{d}}
	\alpha_{d,c} \mathbf{a}_{\rm BS}\ecs{\phi_{d,c},\theta_{d,c}}\eexpo{-\jmath\frac{2\pi}{K}\left( k-1\right) \tau_{d,c}},
\end{equation}
\begin{equation*}
	1 \leq k \leq K, \quad \alpha_{c}\sim\egausd{0}{\sigma_{d,c}^{2}}, \quad \sigma_{d}^{2} = \sum_{c=1}^{C_{d}} \sigma_{d,c}^{2},
\end{equation*}
where $C_{d}$ is the number of clusters, $\tau_{d,c}$ accounts for the delay of the $c$-th cluster measured in samples, $\alpha_{d,c}$ is the channel coefficient for the $c$-th cluster, $\sigma_{d,c}^{2}$ is the average gain of the $c$-th cluster and $\sigma_{d}^{2}$ accounts for the total gain of the channel. In turn, $\mathbf{a}_{\rm BS}\ecs{\phi_{d,c},\theta_{d,c}}$ corresponds to the array steering vector at the BS, and its arguments are the azimuth and elevation angles of arrival, respectively, for the $c$-th cluster. The steering vector for the BS is given by
\begin{equation}\label{eqn:ura1}
	\mathbf{a}_{\rm BS}\ecs{\phi,\theta} = \mathbf{a}_{\rm BS}^{x}\ecs{\phi,\theta} \otimes \mathbf{a}_{\rm BS}^{y}\ecs{\phi,\theta},
\end{equation}
\begin{equation}\label{eqn:ura2}
	\eelem{\mathbf{a}_{\rm BS}^{x}\ecs{\phi,\theta}}{}{b} = 
	\eexpo{\jmath\frac{2\pi}{\lambda} (b-1) d_{H}^{\text{BS}}\esine{\theta}\ecosi{\phi} },
\end{equation}
\begin{equation}\label{eqn:ura3}
	\eelem{\mathbf{a}_{\rm BS}^{y}\ecs{\phi,\theta}}{}{b} = 
	\eexpo{\jmath\frac{2\pi}{\lambda} (b-1) d_{V}^{\text{BS}}\esine{\theta}\esine{\phi} },
\end{equation}
where $1 \leq b \leq B$ is the antenna index and $\lambda$ is the wavelength, as well as $\mathbf{a}_{\rm BS}^{x}(\phi,\theta)$ and $\mathbf{a}_{\rm BS}^{y}(\phi,\theta)$ are the steering vector for the horizontal and vertical arrays, respectively.

The channel impulse response of the BS-RIS link at $k$-th subcarrier can be described as 
\begin{equation}\label{eqn:ref_bs_ris_los}
	\overline{\mathbf{G}}_{e,k}  \triangleq \mu_{e}
	\mathbf{a}_{\rm BS}\ecs{\phi_{e},\theta_{e}}
	\mathbf{a}_{\rm RIS}^{T}\ecs{\varphi_{e},\vartheta_{e}} \eexpo{-\jmath\frac{2\pi}{K}\left( k-1\right) {\tau_{e}}},
\end{equation}
\begin{equation}\label{eqn:ref_bs_ris_nlos}
	\begin{split}
		\tilde{\mathbf{G}}_{e,k}  \triangleq & \sum_{c=1}^{C_{e}}\alpha_{e,c}
		\mathbf{a}_{\rm BS}\ecs{\phi_{e,c},\theta_{e,c}}
		\mathbf{a}_{\rm RIS}^{T}\ecs{\varphi_{e,c},\vartheta_{e,c}} \\
		& \times \eexpo{-\jmath\frac{2\pi}{K}\left( k-1\right) \tau_{e,c}},
	\end{split}
\end{equation}
where $\mu_{e}$ and $\alpha_{e,c}$ are the channel coefficients, $\tau_{e}$ and $\tau_{e,c}$ are the delay measured in samples. Similarly to the BS, $\mathbf{a}_{\rm RIS}\ecs{\phi_{e},\theta_{e}}$, $\mathbf{a}_{\rm RIS}\ecs{\phi_{e,c},\theta_{e,c}}$, $\mathbf{a}_{\rm RIS}\ecs{\varphi_{e},\vartheta_{e}}$ and $\mathbf{a}_{\rm RIS}\ecs{\varphi_{e,c},\vartheta_{e,c}}$ denote the steering vectors for the RIS, and their arguments are the azimuth and elevation angles of arrival and departure, respectively. The expression for the steering vector is the same as described in (\ref{eqn:ura1})-(\ref{eqn:ura3}), replacing respectively ($B_{H}$, $B_{V}$, $d_{H}^{\text{BS}}$, $d_{V}^{\text{BS}}$) by ($M_{H}$, $M_{V}$, $d_{H}^{\text{RIS}}$, $d_{V}^{\text{RIS}}$).

The channel impulse response of the RIS-UE link at $k$-th subcarrier can be described as 
\begin{equation}\label{eqn:ref_ris_ue_los}
	\overline{\mathbf{g}}_{u,k}  \triangleq \mu_{u} 
	\mathbf{a}_{\rm RIS}\ecs{\varphi_{u},\vartheta_{u}}
	\eexpo{-\jmath\frac{2\pi}{K}\left( k-1\right) {\tau_{u}}},
\end{equation}
\begin{equation}\label{eqn:ref_ris_ue_nlos}
	\tilde{\mathbf{g}}_{u,k}  \triangleq \sum_{c=1}^{C_{u}} \alpha_{u,c} 
	\mathbf{a}_{\rm RIS}\ecs{\varphi_{u,c},\vartheta_{u,c}}
	\eexpo{-\jmath\frac{2\pi}{K}\left( k-1\right) \tau_{u,c}},
\end{equation}
where $\mu_{u}$ and $\alpha_{u,c}$ are the channel coefficients, $\tau_{u}$ and $\tau_{u,c}$ account for the delay measured in samples, and the arguments of the steering vectors correspond to the angles of arrival to the RIS.

\subsection{Uplink Transmission}

Given the channel coherence time ($T_{c}$), the UE transmits a frame of $N$ contiguous OFDM symbols of $K$ subcarriers each. In order to avoid the Inter-Symbol and Inter-Carrier Interferences (ISI and ICI), the length $L_{CP}$ of the cyclic prefix must be long enough to absorb the BS-UE direct and reflective paths.

At the BS, the baseband representation of the received signal $\mathbf{y}_{k,n}\ecsizeo{B}$ at the $k$-th subcarrier in the $n$-th OFDM symbol is given by
\begin{equation}\label{eqn:rx_y}
	\mathbf{y}_{k,n} = \mathbf{h}_{k,n}x_{k,n}  + \mathbf{v}_{k,n}, \quad 1\leq k \leq K, \quad 1 \leq n \leq N,
\end{equation}
where $x_{k,n}\in\mathbb{C}$ denotes the complex symbol transmitted from the UE whose power is $\eexpabstwo{x_{k,n}}=P_{x}$, $\mathbf{v}_{k,n}\ecsizeo{B}$ represents the Additive White Gaussian Noise (AWGN) vector which is distributed as $\eelem{\mathbf{v}_{k,n}}{}{b}\sim\egausd{0}{\sigma_{v}^{2}}$, and $\mathbf{h}_{k,n}\ecsizeo{B}$ is the effective channel frequency response between BS and UE, which can be decomposed as
\begin{equation}\label{eqn:model_chan1}
	\mathbf{h}_{k,n} \triangleq \mathbf{h}_{d,k} + \mathbf{h}_{r,k,n} =
	\mathbf{h}_{d,k} + \mathbf{H}_{q,k}\boldsymbol{\psi}_{n},
\end{equation}
\begin{equation*}
	1\leq k \leq K, \quad 1 \leq n \leq N,
\end{equation*}
where $\mathbf{h}_{d,k} \ecsizeo{B}$ is the direct BS-UE channel frequency response at the $k$-th subcarrier, $\mathbf{h}_{r,k,n} \ecsizeo{B}$ corresponds to the reflective BS-UE channel frequency response through the RIS at the $k$-th subcarrier in the $n$-th OFDM symbol. The symbol $\boldsymbol{\psi}_{n}\ecsizeo{M}$ accounts for the phase configurations applied to the RIS at the $n$-th OFDM symbol, which is defined as
\begin{equation}\label{eqn:panel}
	\boldsymbol{\psi}_n \triangleq \begin{bmatrix} \eexpo{\jmath \psi_{n,1}} & \cdots & \eexpo{\jmath\psi_{n,M}} \end{bmatrix}^{T}, \quad 1 \leq n \leq N,
\end{equation}
with $\psi_{n,m}$ for $1\leq m \leq M$ representing the $m$-th phase shift of the RIS. Moreover, $\mathbf{H}_{q,k}\ecsize{B}{M}$ denotes the cascaded channel frequency response at $k$-th subcarrier which is given by
\begin{equation}\label{eqn:cascaded_chan}
	\eelem{\mathbf{H}_{q,k}}{}{b,m} = \eelem{\mathbf{G}_{e,k}}{}{b,m} \eelem{\mathbf{g}_{u,k}}{}{m},
\end{equation}
\begin{equation*}
	1\leq k \leq K, \quad 1 \leq b \leq B, \quad 1 \leq m \leq M.
\end{equation*}

\section{Beam Training for RIS-aided communications}
\label{sec:beamtraining}

Beam training for RIS-aided communications \cite{ris_beam_geo,ris_beam_rui,george5,newr1,newr2} consists of transmitting some reference signals using different phase configurations of a predefined codebook, similarly to the proposed scheme for multi-antenna BS at mm-Wave \cite{beam_chenhao,beam_george}. Each phase configuration will produce a (possibly directive) beam, and hence, the receiver will be able to estimate the received energy for each phase configuration. Finally, the chosen beam corresponds to that phase configuration associated to the highest received energy. Note that the beam training procedure requires a significant amount of time to scan the whole three dimensional space. Therefore, in order to accelerate this process, multi-beam training is proposed, where the panel is split into several sub-panels and each of them is in charge of a portion of the entire space. However, this time reduction comes at the expense of reducing the directivity of the beams, and consequently, UEs far from the BS cannot be detected.

In the following subsection, the average channel gain of an RIS-aided communication system, based on a geometric wideband channel model, is characterized by using the best phase configuration at the RIS for the UE of interest. Then, given this upper-bound and the considered channel model, a codebook of phase configurations will be presented, with a similar approach to \cite{ris_beam_geo,ris_beam_rui,george5,newr1,newr2}. For the sake of simplicity and ease of notation, the described beam training procedure is based on a single beam. Note that the proposed beam training process can be easily extended to the multi-beam case.

\subsection{Channel Gain Enhancement by RIS}

Taking into account (\ref{eqn:rx_y}) and considering that the channel impulse responses, transmitted data symbols and noise are independent random variables, the average received signal energy at the BS for a given RIS phase configuration ($\boldsymbol{\psi}_{n}$) is described by
\begin{equation} \label{eq:rx_energy}
	\begin{split}
		& \eexpnortwo{\mathbf{y}_{k,n}} = \left( \eexpnortwo{\mathbf{h}_{d,k}} + \eexpnortwo{\mathbf{h}_{r,k,n}}\right) P_{x} \\
		& + \sigma_{v}^{2} = \left( \eexpnortwo{\mathbf{h}_{d,k}} + \eexpnortwo{\mathbf{H}_{q,k} \boldsymbol{\psi}_{n}} \right) P_{x} + \sigma_{v}^{2}, \\
	\end{split}
\end{equation}
where $1 \leq n \leq N$ and the expectation is performed over the subcarriers. Since the reflective BS-UE channel gain can be enhanced by manipulating $\boldsymbol{\psi}_{n}$, the optimization problem can be formulated as
\begin{equation} \label{eqn:maxgain}
	\underset{ \boldsymbol{\psi}_{n} }{\text{max}} \quad \eexpnortwo{\mathbf{H}_{q,k} \boldsymbol{\psi}_{n}},\quad \text{s.t.} \quad  \boldsymbol{\psi}_{n} \in \mathcal{D},
\end{equation}
where $\mathcal{D}$ is the codebook of available phase configurations (see, e.g. \cite{Beams_Rahal}, for realistic values for the number of RIS tunable elements). 

By inspecting the cascaded channel ($\mathbf{H}_{q,k}$) given in (\ref{eqn:cascaded_chan}),which is an element-wise product of the BS-RIS and RIS-UE channel frequency responses in (\ref{eqn:ref_bs_ris_all}) and (\ref{eqn:ref_ris_ue_all}), respectively, the RIS is capable to generate a pencil-like sharp beam and focus the energy towards a specific angular direction, the phase configuration capable of maximizing (\ref{eqn:maxgain}) is given by
\begin{equation}\label{eqn:bestbeam}
	\boldsymbol{\psi}^{+} =  \ecs{\mathbf{a}_{\rm RIS}\ecs{\varphi_{e},\vartheta_{e}} \odot \mathbf{a}_{\rm RIS}\ecs{\varphi_{u},\vartheta_{u}}}^{*},
\end{equation}
where $\boldsymbol{\psi}^{+}$ is the best phase configuration of the RIS, and according to \cite{newr3_stat} and \cite{ris_beam_geo}, it is capable of simultaneously pointing to the angular direction of the LoS components of the BS-RIS and RIS-UE links, ($\varphi_{e},\vartheta_{e}$) from (\ref{eqn:ref_bs_ris_los}) and ($\varphi_{u},\vartheta_{u}$) from (\ref{eqn:ref_ris_ue_los}), respectively, which are the strongest taps ($\delta_{e},\delta_{u}>1$). Hence, the best reflective channel can be rewritten as
\begin{equation}\label{eqn:ref_all_mat}
	\begin{split}
		\mathbf{h}_{r,k}^{+} =& \mathbf{H}_{q,k} \boldsymbol{\psi}^{+} \approx \sqrt{L_{e}L_{u}} \mu_{e}\mu_{u} M \mathbf{a}_{\rm BS}\ecs{\phi_{r},\theta_{r}} \\
		& \times \eexpo{-\jmath\frac{2\pi}{K}\left( k-1\right)\ecs{\tau_{e}+\tau_{u}}}.
	\end{split}
\end{equation}
Note that $\mathbf{h}_{r,k}^{+}$ is an upper-bound since the amplitude gain of the LoS component is amplified by the number of passive elements of the panel ($M$) in (\ref{eqn:ref_all_mat}).

\subsection{Phase Configuration Codebook}

Taking into account \cite{ris_beam_geo} and following the analysis given in (\ref{eqn:maxgain})-(\ref{eqn:ref_all_mat}), the codebook $\mathcal{D}$ can be built as
\begin{equation} \label{eqn:codebook}
	\mathcal{D} = \left\lbrace \tilde{\boldsymbol{\psi}}_{1}, \tilde{\boldsymbol{\psi}}_{2}, \ldots, \tilde{\boldsymbol{\psi}}_{N_{CB}}\right\rbrace, \quad \eabsn{\mathcal{D}}{}=N_{CB},
\end{equation}
where $N_{CB}$ is the number of entries of the codebook, and the $i$-th entry of $\mathcal{D}$ is computed as
\begin{equation} \label{eqn:codebooki}
	\tilde{\boldsymbol{\psi}}_{i} = \ecs{\mathbf{a}_{\rm RIS}\ecs{\varphi_{e},\vartheta_{e}} \odot \mathbf{a}_{\rm RIS}\ecs{\varphi_{u,i_{\varphi}},\vartheta_{u,i_{\vartheta}}}}^{*},
\end{equation}
\begin{equation*}
	i = \ecs{i_{\varphi}-1} N_{\varphi} + i_{\vartheta}, 
	\quad 1\leq i_{\varphi}\leq N_{\varphi}, 
	\quad 1\leq i_{\vartheta}\leq N_{\vartheta},
\end{equation*}
with $N_{\varphi}$ and $N_{\vartheta}$ being the number of codewords in the azimuth and elevation dimensions, respectively, and $N_{CB} = N_{\varphi}N_{\vartheta}$. Note that the angles $\varphi_{e}$ and $\vartheta_{e}$ in (\ref{eqn:codebooki}) are considered to be known at the BS since the channel of BS-RIS link is time invariant. On the other hand, the azimuth and elevation angles $\ecs{\varphi_{u,i_{\varphi}},\vartheta_{u,i_{\vartheta}}}$ can be obtained as
\begin{equation} \label{eqn:codephi}
	\varphi_{u,i_{\varphi}} = \ecs{i_{\varphi}-1}\Delta \varphi, \quad 1\leq i_{\varphi}\leq N_{\varphi}, \quad \Delta \varphi N_{\varphi} = \pi,
\end{equation}
\begin{equation} \label{eqn:codetheta}
	\vartheta_{u,i_{\vartheta}} = \frac{\pi}{2} + \ecs{i_{\vartheta}-1}\Delta \vartheta, \quad 1\leq i_{\vartheta}\leq N_{\vartheta}, \quad \Delta \vartheta N_{\vartheta} = \frac{\pi}{2},
\end{equation}
where $\Delta \varphi$ and $\Delta \vartheta$ represent the angle resolution in the azimuth and elevation dimensions.

\section{Proposed NCDS-RIS}
\label{sec:nc-diff}

The beam training process \cite{ris_beam_rui,ris_beam_geo,george5,newr1,newr2,beam_chenhao,beam_george} relies on the transmission of reference signals in order to allow the received energy to be measured. Furthermore, once the best beam of the codebook is chosen, the data symbols are transmitted and received by using the classical CDS, which also relies on reference signals. Consequently, in order to successfully deploy an RIS-empowered communication system, the transmission of a significant amount of overhead is required, which implies a reduction of the system efficiency, especially for high mobility scenarios where the required amount of pilot symbols is much larger.

The proposal NCDS-RIS, which consists of exploiting NCDS based on differential modulation \cite{Ana2015,Kun2019,Kun2020,KunJFET} at RIS-aided communications during and after the beam training process is presented in this section (see Fig. \ref{fig:training}). Note that the whole beam training procedure is centralized at the BS and it is transparent to all UEs of the cell. During the beam training process, the effective BS-UE channel impulse response ($\mathbf{h}_{k,n}$) is extremely time-varying due to the continuous testing of different phase configurations of the codebook at the RIS. However, the differentially encoded data are robust to channel variations. They can be transmitted and demodulated without the need of any channel estimation, and at the same time, their received power can be measured. In the second stage, once the best codeword is chosen and configured at the RIS panel, we will show that the advantage of avoiding the transmission of additional reference signals in the NCDS is still interesting for practical scenarios, especially when the channel coherence time ($T_{c}$) is not excessively large. Consequently, the proposed scheme not only is capable of avoiding the overhead produced by the undesirable reference signals, but it also does not require any network information.

The $N$ OFDM symbols transmitted during the channel coherence time ($T_{c}$) are split into two stages (see Fig. \ref{fig:training}). In the first stage, $N_{l}$ OFDM symbols are transmitted from the UE to the BS through the direct link, and at the same time, the panel will test different phase configurations given by the codebook. In the second stage, the remaining $N_{h}$ OFDM symbols are transmitted by mainly using the enhanced reflected link, thanks to the selection of the best phase configuration of the codebook for the RIS. Consequently, at the second stage, the communication link via RIS provides a higher gain than the direct link, and higher order modulations can be exploited, producing a better throughput.

\eAddFig{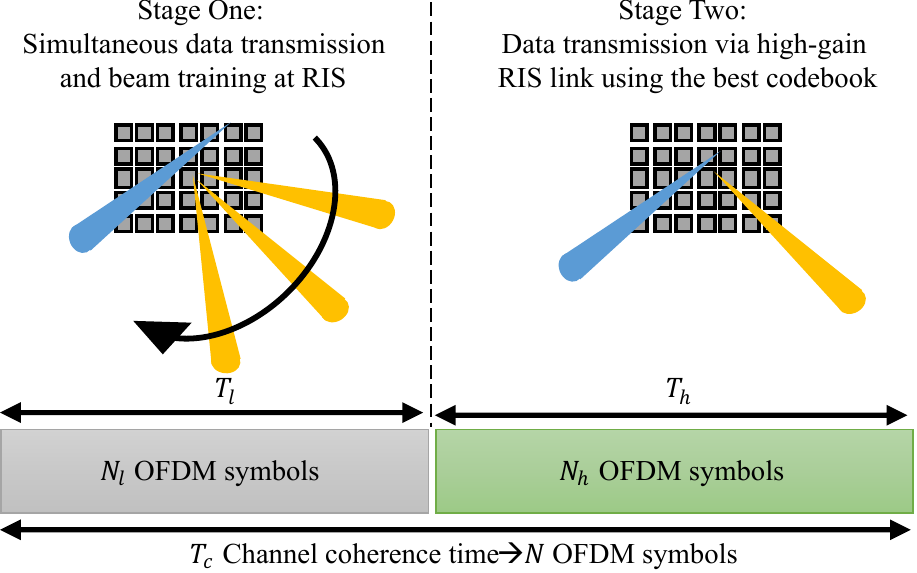}{1}{fig:training}{A frame of $N$ OFDM symbols is transmitted by two stages within the channel coherence time ($T_{c}$). In the first stage, beam training is executed, and at the same time, non-coherent transmission is performed over the BS-UE direct link. In the second stage, the best phase configuration is loaded to the RIS and non-coherent data symbols are transmitted over the enhanced reflective BS-UE link via RIS.}

\subsection{Stage One: Simultaneous Data Transmission and Beam Training at the RIS}

The beam training is performed over the first consecutive $N_{l}$ OFDM symbols, where each entry of the codebook is configured at the RIS for, at least, the duration of one OFDM symbol ($N_{l}\geq N_{CB}$). At this stage, even though all the channel links remain invariant during the channel coherence time ($T_{c}$), the effective BS-UE channel ($\mathbf{h}_{k,n}$), given in (\ref{eqn:model_chan1}), changes from one to the following OFDM symbol, ensuring that the effective coherence time is larger than one OFDM symbol. This effect is produced by the reflective link ($\mathbf{h}_{r,k,n}$), which is constantly varying as a consequence of using different codewords ($\boldsymbol{\psi}_{n}$) at each OFDM symbol. In order to be able to transmit data during this training period, the differential modulation is performed between consecutive subcarriers. In particular, we deploy the Frequency-Domain Scheme (FDS) of \cite{Kun2019}.

At the UE, the data symbols are differentially encoded in the frequency domain before their transmission as follows:
\begin{equation} \label{eqn:diff_enc}
	x_{k,n}  =
	\left\{\begin{array}{@{}cl}
		p_{k,n},  & k=1 \\
		x_{k-1,n}p_{k,n},  & k=2 \\
		x_{k-1,n} s_{k,n}, & 3\leq k \leq K\\
	\end{array}
	\right., \quad 1 \leq n \leq N.
\end{equation}
where $s_{k,n}$ denotes the complex symbol to be transmitted at the $k$-th subcarrier and the $n$-th OFDM symbol, that belongs to a $Q_{l}$-PSK constellation and its power is normalized (i.e., $\eabsn{s_{k,n}}{2}=1$). The constellation size at this stage may be small since the direct BS-UE channel link does not have a LoS component and the path-loss may be high. In (\ref{eqn:diff_enc}), $p_{1,n}$ and $p_{2,n}$ are two reference symbols. Before data transmission, the power of differential symbols $x_{k}^{n}$ is scaled according to $P_{x}$.

According to \cite{Kun2019,KunJFET}, the differential modulation performed at FDS is chosen when the channel coherence time is as small as one OFDM symbol period ($T_{c} \approx \ecs{K+L_{CP}}/\ecs{K\Delta f}$ with $\Delta f$ being the subcarrier spacing measured in Hz). Here, the data symbols are differentially encoded in the contiguous frequency resources of each OFDM symbol. However, this scheme requires two reference symbols ($p_{1,n}$ and $p_{2,n}$), namely the first one to estimate and compensate the residual differential phase component ($\zeta_{l,n}$) produced by the multi-path channel at the $n$-th OFDM symbol at the first stage, and the second one for performing the differential demodulation. Nevertheless this overhead can be neglected for broadband systems, when $K$ is very large.  

Given the received signal (\ref{eqn:rx_y}), the BS performs the following differential decoding:
\begin{equation} \label{eqn:diff_dec}
	\begin{split}
		z_{k,n} & = \frac{1}{B} \eherm{\mathbf{y}_{k-1,n}}\mathbf{y}_{k,n} \eexpo{-\jmath\widehat{\zeta}_{l,n}} \\
		& = \frac{1}{B}\eexpo{-\jmath\widehat{\zeta}_{l,n}}\sum_{i=1}^{4} I_{i}, \\
	\end{split}
\end{equation}
\begin{equation} \label{eqn:diff_dec1}
	I_{1} = \eherm{\mathbf{h}_{k-1,n}}\mathbf{h}_{k,n} s_{k,n}, \quad I_{2} = \eherm{\mathbf{h}_{k-1,n}x_{k-1,n}}\mathbf{v}_{k,n},
\end{equation}
\begin{equation} \label{eqn:diff_dec3}
	I_{3} = \eherm{\mathbf{v}_{k-1,n}}\mathbf{h}_{k,n} x_{k,n}, \quad I_{4} = \eherm{\mathbf{v}_{k-1,n}} \mathbf{v}_{k,n},
\end{equation}
\begin{equation*}
	3 \leq k \leq K, \quad 1 \leq n \leq N,
\end{equation*}
where $I_{1}$ includes the useful symbol $s_{k,n}$, $I_{2}$ and $I_{3}$ represent the cross-interference terms produced by the noise and the received differential symbol in two time instants, while $I_{4}$ is exclusively produced by the product of the noise in two contiguous subcarriers. Moreover, the residual phase component at the $n$-th OFDM symbol can be estimated as
\begin{equation} \label{eqn:diff_phase}
	\widehat{\zeta}_{l,n} = \measuredangle \ecs{\eherm{\mathbf{y}_{1,n}}\mathbf{y}_{2,n}} - \measuredangle \ecs{\frac{p_{2,n}}{p_{1,n}}}, \quad 1 \leq n \leq N.
\end{equation}

While this transmission occurs, the beam training process is being executed at the RIS, where a different phase configuration of the codebook is configured for each group of $N_{l}$ consecutive OFDM symbols, and the BS is measuring the received power in the $n$-OFDM symbol as follows:
\begin{equation} \label{eqn:meas_pow}
	P_{y,n} = \frac{1}{K}\sum_{k=1}^{k} \eexpnortwo{\mathbf{y}_{k,n}}, \quad 1 \leq n \leq N_{l}.
\end{equation}
After testing all entries of the codebook, the BS will choose that phase configuration corresponding to the highest measured received power (\ref{eqn:meas_pow}). Hence, the gain of the reflective channel is significantly enhanced by the RIS (\ref{eqn:bestbeam})-(\ref{eqn:ref_all_mat}). Besides, note that the differential data transmission has the benefit that it can be combined with any beam training procedure (single-beam, multi-beam, hierarchical, etc.), since it is independent of the particular choice of codebook, and any anti-jamming processing.

\subsection{Stage Two: Data Transmission via Reflective RIS Link}

Once the best phase configuration is chosen in the previous stage, the effective BS-UE channel remains constant for $N_{h}$ OFDM symbols ($\mathbf{h}_{k,n}=\mathbf{h}_{k}$, $1 \leq n \leq N_{h}$). Consequently, the differential modulation can be performed by using the Mixed Domain Scheme (MDS) \cite{KunJFET}, where the differential data can be simultaneously encoded in both time and frequency domains. The main benefit of this scheme is that only two reference signals are required for transmission of a total of $N_{h}$ OFDM symbols ($KN_{h}$ resources), reducing further the overhead as compared to the FDS.

Firstly, the data symbols are differentially encoded as
\begin{equation} \label{eq:mds}
	\tilde{x}_{i}  =
	\left\{\begin{array}{@{}cl}
		p_{i},  & i=1 \\
		\tilde{x}_{i-1}p_{i},  & i=2 \\
		\tilde{x}_{i-1}s_{i}, & 3 \leq i \leq KN_{h} \
	\end{array}
	\right.,
\end{equation}
where the $i$ denotes the resource index. Then, the differential symbols $\tilde{x}_{i}$ are allocated to the two-dimensional resource grid as
\begin{equation} \label{eq:mapping}
	x_{k,n} = \tilde{x}_{i} \mid (k,n)=f(i), \quad 1\leq i \leq KN_{h},
\end{equation}
where $f(\cdot)$ is the resource mapping policy function. A possible mapping policy is given in \cite{KunJFET}, which mainly follows the FDS, except for the edge subcarriers of the block, that follow a Time-Domain Scheme (TDS). The latter consists on performing the differential encoding using resources of the same subcarrier at two consecutive OFDM symbols.

Similarly to the previous stage, the residual phase compensation is also required, except for those differential symbols following the TDS. The residual phase component for the second stage ($\zeta_{h}$) can be estimated by using only the first two subcarriers of the first OFDM symbol as follows:
\begin{equation} \label{eqn:diff_phase2}
	\begin{split}
		\widehat{\zeta}_{h} = \measuredangle \ecs{\eherm{\mathbf{y}_{1,1}}\mathbf{y}_{2,1}} - \measuredangle \ecs{\frac{p_{2,1}}{p_{1,1}}}.
	\end{split}
\end{equation}

\section{Analysis of the SINR}
\label{sec:sinr}

In this section, the SINR performance of NCDS-RIS is analyzed. According to \cite{Kun2019,Kun2020,KunJFET}, the residual phase component can be considered perfectly estimated and compensated, since the degradation produced by the residual phase value is negligible after estimation and compensation.

Taking into account (\ref{eqn:diff_dec1}) and (\ref{eqn:diff_dec3}), the undesirable effects produced by the self-interference and noise terms can be characterized as 
\begin{equation} \label{eqn:evm}
	\begin{split}
		& \eexpabstwo{\ecs{\beta_{d}^{2}+\beta_{r}^{2}}  s_{k,n}-z_{k,n}} = \ecs{\beta_{d}^{2}+\beta_{r}^{2}}^{2} P_{x}^{2} + \\
		& \eexpabstwo{z_{k,n}} -2\ecs{\beta_{d}^{2}+\beta_{r}^{2}}\ereal{\eexp{\eherm{s_{k,n}}z_{k,n}}},
	\end{split}
\end{equation}
\begin{equation} \label{eqn:beta_d}
	\beta_{d}^{2} = L_{d} \sigma_{d}^{2}, \quad \beta_{r}^{2} \leq L_{e}L_{u} \mu_{e}^{2} \mu_{u}^{2}M^{2},
\end{equation}
where $\beta_{d}^{2}$ and $\beta_{r}^{2}$ are the channel gains of the direct and reflective channel links, respectively, which is taking into account both the large and small-scale effects of the channel. The expectation is performed over the subcarriers and OFDM symbols. Note that the channel gain of the reflective BS-UE link is upper-bounded by using the best phase configuration given in (\ref{eqn:bestbeam}), which is capable of perfectly matching with the arrival and departure angles of the LoS components of the BS-RIS and RIS-UE channel links. 

Following the derivations given in \cite{Ana2015,Kun2019,Kun2020,KunJFET}, the four terms given in (\ref{eqn:diff_dec1}) and (\ref{eqn:diff_dec3}) are statistically independent since the channel frequency response, noise, and symbols are independent random variables, and the noise samples between two time instants are also independent. Hence, the two last terms in (\ref{eqn:evm}) can be simplified as
\begin{equation} \label{eqn:victor1}
	\eexpabstwo{z_{k,n}} = \frac{1}{B^2}\sum_{i=1}^{4}\eexpabstwo{I_{i}},
\end{equation}
\begin{equation} \label{eqn:victor2}
	\eexp{\eherm{s_{k,n}}z_{k,n}} = \frac{1}{B}\eexp{\eherm{s_{k,n}}I_{1}}.
\end{equation}
After performing some manipulations given in Appendix \ref{appendix:geo}, the different interference and noise terms are given by
\begin{equation} \label{eqn:sI1}
	\eexp{\eherm{s_{k,n}}I_{1}} = \ecs{\beta_{d}^{2} +\beta_{r}^{2}} B P_{x}^{2},
\end{equation}
\begin{equation} \label{eqn:power_i1}
	\eexpabstwo{I_{1}} = \ecs{2\beta_{d}^{4}+\beta_{r}^{4}+4\beta_{d}^{2}\beta_{r}^{2}} B^{2} P_{x}^{2},
\end{equation}
\begin{equation} \label{eqn:power_i23}
	\eexpabstwo{I_{2}} = \eexpabstwo{I_{3}} = \sigma_{v}^{2} \ecs{\beta_{d}^{2} +\beta_{r}^{2}} B P_{x},
\end{equation}
\begin{equation} \label{eqn:power_i4}
	\eexpabstwo{I_{4}} = B \sigma_{v}^{4}.
\end{equation}
Hence, the SINR of the proposed NCDS-RIS assuming a geometrical wideband channel is given by
\begin{equation} \label{eqn:sinr}
	\rho = \einv{2\frac{\sigma_{v}^{2}}{B\ecs{\beta_{d}^{2}+\beta_{r}^{2}}P_{x}}+\frac{2\beta_{d}^{2}\beta_{r}^{2}+\beta_{d}^{4}+\sigma_{v}^{4}/B}{\ecs{\beta_{d}^{2}+\beta_{r}^{2}}^{2}P_{x}^{2}}},
\end{equation}
where the SINR can be enhanced by increasing the transmit power ($P_{x}$), the number of antennas at the BS ($B$), and/or the channel gain of the direct and reflective links ($\beta_{d}^{2}$, $\beta_{r}^{2}$, respectively). However, the first term of the second fraction is saturating the quality of the link due to the presence of the direct link ($\beta_{d}^{2}$). 

The SINR given in (\ref{eqn:sinr}) will be particularized for two extreme but illustrative scenarios. The first scenario corresponds to the performance obtained by using exclusively the direct link between BS-UE, where the reflective link is absent. On the contrary, the second case only takes into account the reflective link, assuming that the direct link is negligible. 

\subsection{Direct Link Only (Lower-Bound)}

Before choosing a proper phase configuration of the codebook for the RIS, the gain of the direct link is typically much higher than the reflected one by the RIS ($\beta_{d}^{2} >> \beta_{r}^{2}$). Given this situation, the data transmission is performed over mainly the direct link, and therefore, the SINR given in (\ref{eqn:sinr}) can be simplified as
\begin{equation} \label{eqn:sinr_d}
	\rho_{d} = \einv{1 + 2\frac{\sigma_{v}^{2}}{B\beta_{d}^{2}P_{x}}+\frac{\sigma_{v}^{4}}{B\beta_{d}^{4}P_{x}^{2}}},
\end{equation}
where the quality of the link is upper-bounded by the first term, pointing out that the performance of the proposed system under a geometric wideband channel with NLoS paths is limited.

\subsection{Reflective Link Only (Asymptotic Case)}

Assuming the hypothetical case that the RIS size is very large ($M\uparrow$), the direct channel can be neglected since the gain of the reflected channel by RIS is significantly high ($\beta_{d}^{2} << \beta_{r}^{2}$), and hence, the SINR given in (\ref{eqn:sinr}) can be simplified as
\begin{equation} \label{eqn:sinr_r}
	\rho_{r} = \einv{\frac{2\sigma_{v}^{2}+\sigma_{v}^{4}}{B\beta_{r}^{2}P_{x}}+\frac{2\beta_{d}^{2}}{\beta_{r}^{2}P_{x}^{2}}} = \frac{B\beta_{r}^{2}P_{x}}{2\sigma_{v}^{2}+\sigma_{v}^{4}}.
\end{equation}
Assuming that $\sigma_{v}^{4} << \sigma_{v}^{2}$ and making use of the best phase configuration given in (\ref{eqn:bestbeam}), (\ref{eqn:sinr_r}) can be approximated as
\begin{equation} \label{eqn:sinr_r2}
	\rho_{r} \approx \frac{B L_{e}L_{u} \mu_{e}^{2} \mu_{u}^{2}M^{2}P_{x}}{2\sigma_{v}^{2}},
\end{equation}
where it is noted that the SINR can be approximated by a linear function, and the performance is directly scaled with the reflective channel gain via the RIS, transmit power, and the number of antennas at the BS. Additionally, note that the constant value at the denominator corresponds to the typical 3 dB of performance loss of NCDS as compared to CDS. However, the classical CDS requires a significant amount of reference signals in order to track the variations of the channel, especially for high mobility scenarios, and hence, the overall throughput is significantly reduced with NCDS\cite{Manu2020}.

\section{Throughput and Complexity Comparison}\label{sec:efficiency}

In this section, a comparison in terms of efficiency and complexity among the proposed NCDS-RIS and the baseline cases are given, in order to show the superiority of the NCDS-RIS.

For the first stage of beam training, two baseline approaches are taken into account for comparison purposes. The transmitted symbols can be either exclusively Reference Signals (RS) \cite{ris_beam_rui}, or the traditional CDS (pilot and data symbols). Therefore, the main difference between these two cases is that CDS is also able to transmit some information. The chosen combiner for this stage is Maximum Ratio Combining (MRC) since the channel gain of the direct link is not very high. 

For the second stage, only the classical CDS is considered. According to \cite{ris_beam_geo}, the first OFDM symbol of this stage is exclusively employed for channel estimation, while the $N_{h}-1$ OFDM symbols left are used for data transmission. Unlike the first stage, the chosen combiner is Zero-Forcing (ZF) since the channel gain of the reflected link is enhanced by the RIS, and for high-SINR ZF is superior to MRC.

\subsection{Throughput Comparison}

For a typical packet-based transmission, the total throughput of the system can be defined as
\begin{equation}\label{eqn:tr_t}
	R =  R_{l} + R_{h} \left[ \text{packet/s}\right] ,
\end{equation}
where $R_{l}$ and $R_{h}$ refer to the throughput of the first and second stages, respectively. Similarly to \cite{Manu2020}, the throughput of each stage can be found as 
\begin{equation} \label{eqn:thput_c}
	R_{l}= \eta_{l}\frac{\Delta f K}{L_{P}}\left( 1-P_{e,l} \right)^{L_{P}} \log_2(Q_{l}),
\end{equation}
\begin{equation} \label{eqn:thput_n}
	R_{h}= \eta_{h}\frac{ \Delta f K}{L_{P}}\left( 1-P_{e,h} \right)^{L_{P}} \log_2(Q_{h}),
\end{equation}
where $P_{e,l}$ and $P_{e,h}$ are the BER for the two stages, $Q_{l}$ and $Q_{h}$ are the constellation sizes of each stage, $L_{P}$ denotes the number of bits in one packet, and $\eta_{l}$ and $\eta_{h}$ are the efficiency of the system for the two stages, taking into account the overhead produced by the transmission of the reference signals. 

The efficiency of the first stage is given by
\begin{equation} \label{eqn:eff_stage_one}
	\eta_{l}=\frac{N_{l}\ecs{K-K_{p}}}{NK},
\end{equation}
where $K_{p}$ is the number of reference symbols at each OFDM symbol. For the CDS scenario, the number of pilot symbols is typically a portion of the total amount of subcarriers ($K_{p} < K$). On the other hand, the pilot-based approach and the proposed NCDS-RIS are two particular cases of (\ref{eqn:eff_stage_one}), where the number of reference signals is fixed to $K_{p}=K$ and $K_{p}=2$, respectively.

After the best phase configuration is chosen, the data transmission is enhanced by the RIS. The efficiency of this second stage for the CDS case is
\begin{equation} \label{eqn:eff_stage_two}
	\eta_{h}^{\rm CDS}=\frac{N_{h}-1}{N},
\end{equation}
where only one OFDM symbol is used for channel estimation. On the other hand, the efficiency of the second stage for the NCDS is
\begin{equation} \label{eqn:eff_stage_two2}
	\eta_{h}^{\rm NCDS}=\frac{N_{h}K-2}{NK},
\end{equation}
where only two reference symbols placed at the first OFDM symbols are required out of $N_{h}$ OFDM symbols. 

Taking into account (\ref{eqn:tr_t})-(\ref{eqn:eff_stage_two2}), the proposed NCDS-RIS for both stages is capable of outperforming the classical CDS, especially for high mobility scenarios, since it does not require to transmit a large amount of reference signals, and its efficiency is not degraded. 

\begin{table}[!t]
	\centering
	\caption{Complexity Comparison between the Proposed NCDS and the Considered Baseline RS and CDS.}
	\label{tab:complexity}	
	\begin{tabular}{|c|c|c|c|}
		\hline
		& \multicolumn{1}{c|}{\textbf{RS}} & \textbf{CDS}               & \textbf{NCDS} \\ \hline
		\textbf{Stage One} & $0$                  & $N_{l}\ecs{BK_{p}+B^{2}\ecs{K-K_{p}}+C_{I}}$ & $2\ecs{K-1}N_{l}$           \\ \hline
		\textbf{Stage Two} &  - & $BK\ecs{\ecs{B^{2}+1}+B\ecs{N_{h}-1}}$                    & $2\ecs{KN_{h}-1}$           \\ \hline
	\end{tabular}
\end{table}

\subsection{Complexity Comparison}
The complexity evaluation is performed accounting for the number of required complex product operations for each case, which are summarized in Table \ref{tab:complexity}. Note that the operations needed for measuring the energy in order to determine the best beam are not considered in the stage one for this comparison, since they are the same for all approaches.

For the first stage, RS does not require any additional operations due to the fact that it does not transmit any data symbols. On the contrary, CDS is transmitting both data and pilot symbols, and hence, it has to estimate the channel and perform a MRC equalization for each OFDM symbol. Note that the computation of the MRC combiner is neglected, since it is obtained by performing a complex conjugate of the estimated channel. Additionally, the channel estimates at the pilot symbol resources must be interpolated to obtain values for the data resources. The complexity of this interpolation is given by $C_{I}$, and this value depends on the chosen algorithm \cite{chan_interp1,chan_interp2}. However, the proposed NCDS only performs the differential encoding and decoding $K-1$ times for each OFDM symbol. 

For the second stage, the classical CDS performs the channel estimation and computation of the ZF combiner at the first OFDM symbol, which requires a matrix inversion. Then, the remaining OFDM data symbols are equalized. Again, the NCDS only requires the differential encoding and decoding operations.

The proposed NCDS-RIS has less complexity in terms of complex products as compared to the traditional CDS, since channel estimation, computation of the combiner, and the equalization operations are not required. Moreover, the classical CDS not only requires more computations, but it also needs more memory to store the channel estimation samples and the computed combiners. Consequently, the proposed NCDS-RIS system is cost-effective and/or able to reduce the delay of the communication system. 

\section{Performance Evaluation Results}\label{sec:num_res}
In this section, several numerical results are provided in order to show the performance of the proposed NCDS-RIS, as compared to the considered baseline cases presented in the previous section, and the accuracy of the analytical results.

For a more realistic performance evaluation, all the links (BS-UE, BS-RIS and RIS-UE) are generated according to the 5G standard \cite{nr-901}. The power-delay profile follows an exponential distribution whose standard deviation is the DS; the azimuth angles of arrival/departure are modeled by a wrapped Gaussian distribution which is characterized by the Azimuth AS of Arrival and Departure (ASA and ASD); and the zenith angles of arrival/departure are modeled by a Laplacian distribution, also characterized by the Zenith angular Spread of Arrival and Departure (ZSA and ZSD). The channel propagation model adopted for the simulation results corresponds to the 3GPP factory scenario of size (60m,120m,3m), in particular, a low angular spread case is chosen. Moreover, a summary of the simulation parameters is provided in Table \ref{tab:simparam}, the location of each network node is given by the Cartesian coordinates $(x,y,z)$ measured in meters, and $f_{c}$ denotes the carrier frequency. The distance between any two contiguous elements of the BS and RIS is set to half wavelength ($d_{H}^{\text{BS}} = d_{V}^{\text{BS}} = d_{H}^{\text{RIS}} = d_{V}^{\text{BRIS}} = \lambda / 2$).

\begin{table}[!t]
	\centering
	\caption{Simulation Parameters}
	\label{tab:simparam}
	\begin{tabular}{|c|c|c|c|c|c|}
		\hline
		\textbf{BS location}            & (0,0,3)   & $\mathbf{f_{c}}$      & 3.5 GHz & \textbf{ASD} & $7^{\rm o}$  \\ \hline
		\textbf{RIS location}           & (10,0,3)  & $\mathbf{\Delta f}$   & 30 KHz  & \textbf{ASA} & $12^{\rm o}$ \\ \hline
		\textbf{UE location}            & (10,12,1) & $\mathbf{L_{d}}$      & -86 dB  & \textbf{ZSD} & $15^{\rm o}$ \\ \hline
		\textbf{Mod. NCDS}              & 4, 16-PSK & $\mathbf{L_{u}}$      & -60 dB  & \textbf{ZSA} & $20^{\rm o}$ \\ \hline
		\textbf{Mod. CDS}               & 4, 16-QAM & $\mathbf{L_{e}}$      & -62 dB  & $\mathbf{K}$ & 1024         \\ \hline
		$\mathbf{C_{d}}$                & 20        & $\bm{\sigma_{v}^{2}}$ & -90 dBW & $\mathbf{M}$ & 64           \\ \hline
		$\mathbf{C_{e}}$, $\mathbf{C_{u}}$ & 10        & $\mathbf{B}$          & 16      & $\mathbf{N}$ & 1000         \\ \hline
	\end{tabular}
\end{table}

The number of codewords to be tested is set to the number of OFDM symbols at the first stage ($N_{CB} = N_{l}$). Additionally, in order to show the quality of the phase configuration based on codeword selection, the performance provided by the best configuration given in (\ref{eqn:bestbeam}) is also shown, in order to see the difference between the realistic case based on the codebook and its upper-bound.

\subsection{Verification of the SINR Analysis for the Proposed NCDS-RIS}
Figure~\ref{fig:sinr} illustrates the SINR performance, as a function of the UE transmit power $P_{x}$ in dBW, of the proposed NCDS-RIS for the direct BS-UE link and effective BS-UE link made up by the direct and reflective link enhanced by the RIS. As clearly shown, the reflective link significantly outperforms the direct link by approximately $29$-$33$ dB. On the other hand, the performance of the phase configuration based on codeword selection is slightly worse than the best one, and this difference can be reduced further as the number of entries of the codebook is increased ($N_{CB}\uparrow$), at the expense of increasing the beam training period ($N_{l}$). It is also shown in this figure that the SINR analysis given in (\ref{eqn:sinr}) and (\ref{eqn:sinr_d}) (plotted with black dotted lines) accurately characterizes the RIS-empowered system performance under geometric wideband channel models.

\subsection{BER comparison}

The BER comparison between the classical CDS and the proposed NCDS-RIS signaling is depicted in Fig.~\ref{fig:ber}. For the stage one, the chosen modulation is $4$-QAM and $4$-PSK for the CDS and NCDS, respectively, and the selected equalization technique for CDS is MRC. For the second stage, the chosen modulation is $16$-QAM and $16$-PSK for the CDS and NCDS, respectively, and the selected equalization technique for CDS is ZF. Additionally, for the particular case of CDS, the particular case of perfect channel estimation (PCE) is also plotted in order to show the lower bound. Even though the CDS is able to outperform the NCDS in terms of BER, this result does not account for the time/frequency and energy resources required for the transmission of reference signals (efficiency) and the complexity described in Section \ref{sec:efficiency}. Moreover, the degradation produced by the CDS is higher than NCDS for the particular case of codeword selection. The reason behind this behavior is due to the fact that CDS requires accurate channel estimates, especially for the particular case of ZF criterion. Otherwise, the computed equalizers will enhance the noise.

\eAddFig{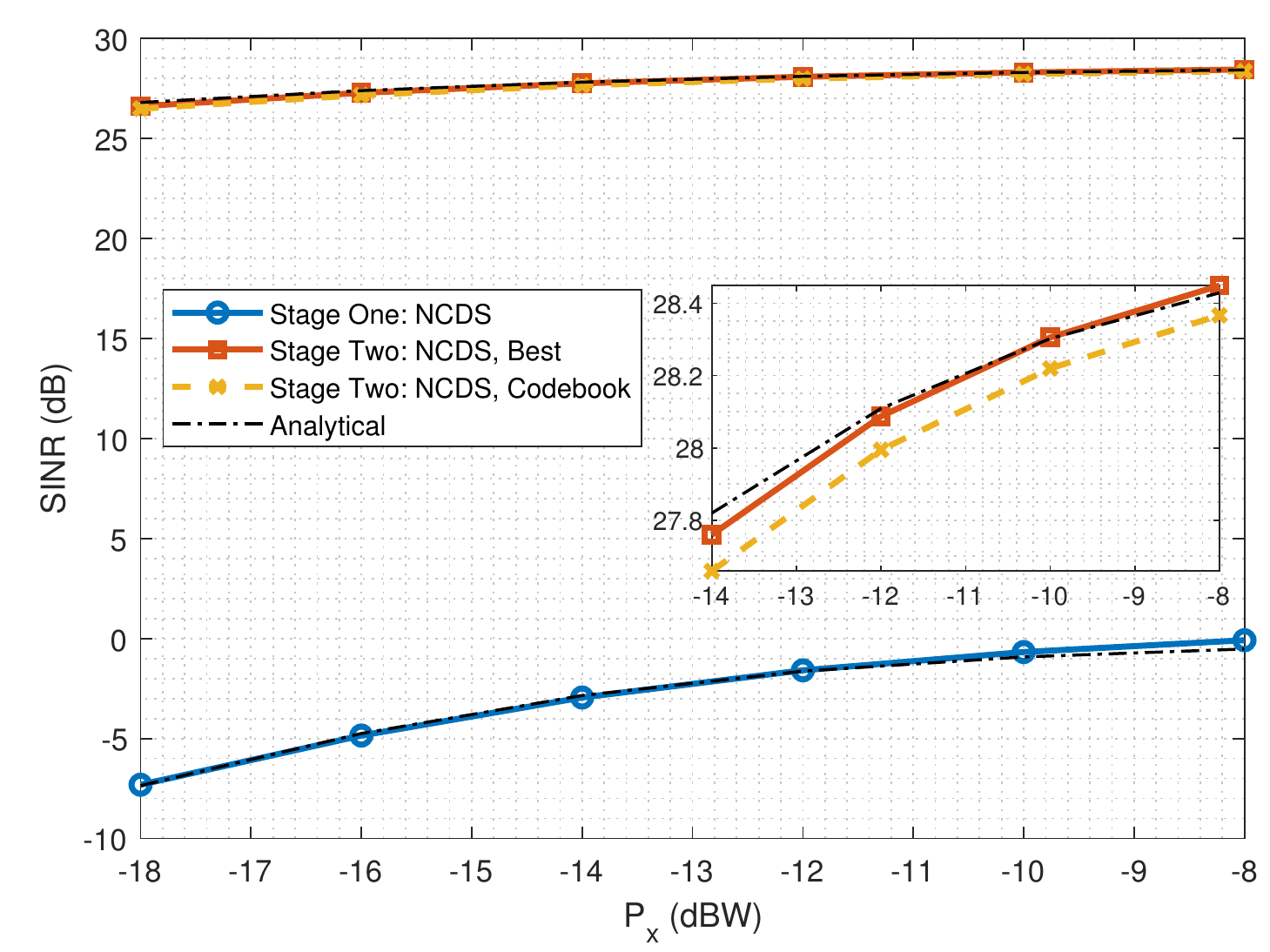}{1}{fig:sinr}{Analytical SINR verification for NCDS at both stages.}
\eAddFig{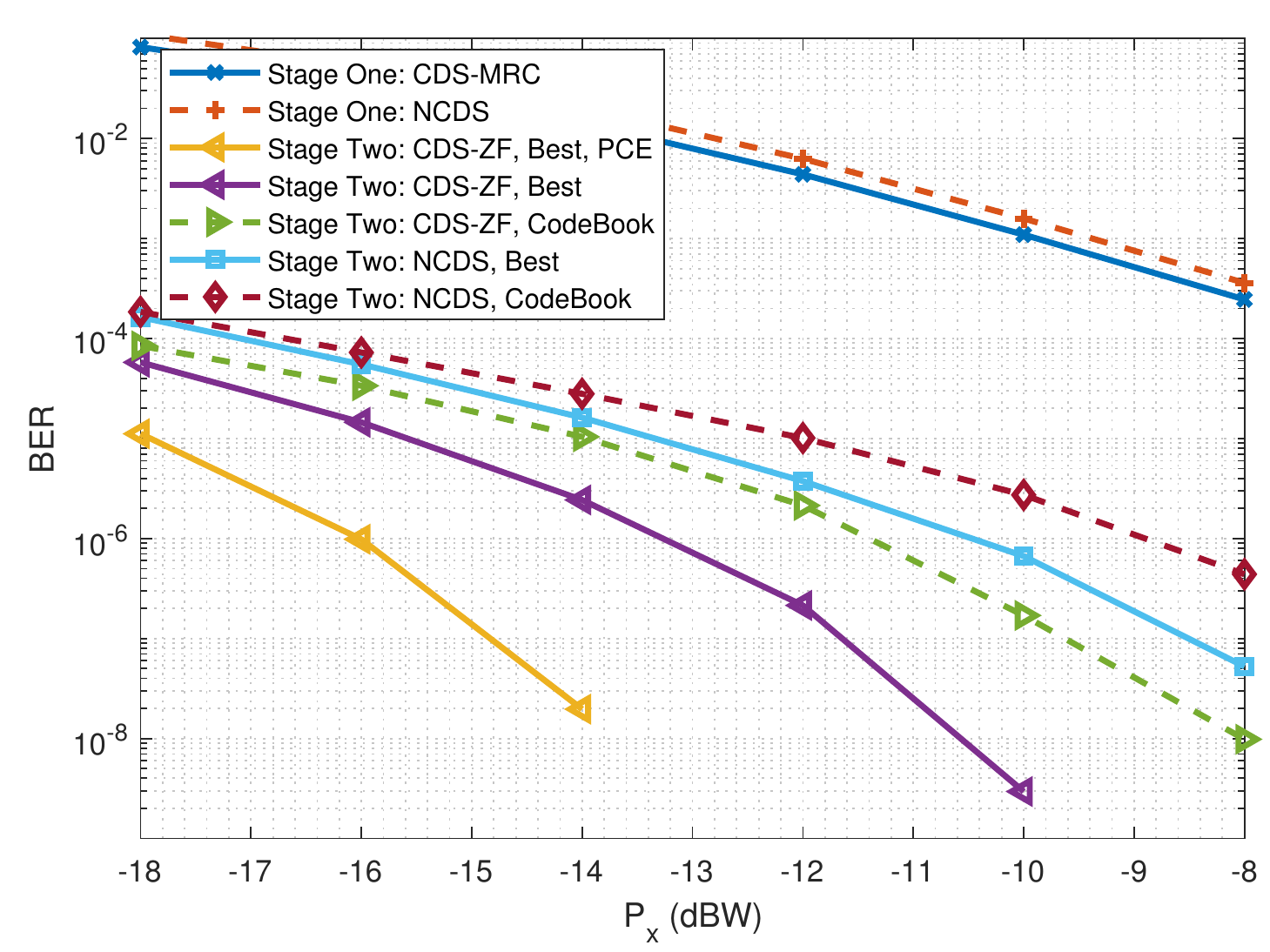}{1}{fig:ber}{BER comparison between CDS and NCDS for both stages.}

\subsection{Throughput Comparison between CDS and NCDS}

Finally, a throughput comparison is shown taking into account the configurations described in Section \ref{sec:efficiency}. The constellation sizes are $Q_{l}=4$ and $Q_{h}=16$, the packet length is set to $L_{P} = 20$, and for the particular case of CDS at the stage one, the ratio of pilot symbols transmitted at each OFDM symbol is set to $K/K_{p} = 3$.

In Table \ref{tab:throughput}, the throughput is evaluated for both stages at different values of speed for the UE, which is equivalent to evaluating the performance at several values of channel coherence times ($T_{c}$) and their corresponding lengths ($N$), taking into account the OFDM numerology given in Table \ref{tab:simparam}. On the one hand, the NCDS is always outperforming the CDS in the first stage, as a result of avoiding the reference signals in order to track the strongly time-varying channels produced by the RIS. Moreover, the throughput of the CDS is slightly lower than the NCDS in the second stage, since the additional OFDM symbol spent for channel estimation becomes negligible. On the other hand, comparing the throughput produced by the two stages, the proposed NCDS-RIS scheme contributes more to the total throughput ($R$) as compared to the CDS, especially  when the channel coherence time or the number of total OFDM symbols is not large. For example, for the case of $7.3$ m/s ($N=N_{l}$), the NCDS can double the throughput.

\begin{table}[!t]
	\centering
	\caption{Throughput comparison between CDS and NCDS at each stage for $P_{x}=-8$ db in [$10^{6}$ Packets/s]. }
	\label{tab:throughput}
	\begin{tabular}{|c|cc|cc|cc|cc|}
		\hline
		& \multicolumn{2}{c|}{\begin{tabular}[c]{@{}c@{}}\textbf{7.3 m/s}\\ $N = N_{l}$\end{tabular}} & \multicolumn{2}{c|}{\begin{tabular}[c]{@{}c@{}}\textbf{4.8 m/s}\\ $N = 1.5N_{l}$\end{tabular}} & \multicolumn{2}{c|}{\begin{tabular}[c]{@{}c@{}}\textbf{3.6 m/s}\\ $N = 2N_{l}$\end{tabular}} & \multicolumn{2}{c|}{\begin{tabular}[c]{@{}c@{}}\textbf{2.4 m/s}\\ $N = 3N_{l}$\end{tabular}} \\ \hline
		\textbf{Stage} & \multicolumn{1}{c|}{One}                        & Two                       & \multicolumn{1}{c|}{One}                          & Two                         & \multicolumn{1}{c|}{One}                         & Two                        & \multicolumn{1}{c|}{One}                         & Two                         \\ \hline
		\textbf{CDS}   & \multicolumn{1}{c|}{2.04}                       & 0                         & \multicolumn{1}{c|}{1.36}                         & 2.04                        & \multicolumn{1}{c|}{1.02}                        & 3.06                       & \multicolumn{1}{c|}{0.68}                        & 4.09                        \\ \hline
		\textbf{NCDS}  & \multicolumn{1}{c|}{3.05}                       & 0                         & \multicolumn{1}{c|}{2.04}                         & 2.05                        & \multicolumn{1}{c|}{1.53}                        & 3.07                       & \multicolumn{1}{c|}{1.02}                        & 4.1                         \\ \hline
	\end{tabular}
\end{table}

In Fig. \ref{fig:throughput}, the total throughput ($R$) is shown for the different baseline cases described in the previous section. The two bottom curves represent the total throughput for the case of exclusively transmitting reference signals at the beam-training case, while both CDS and NCDS are exploited at the second stage. It can be noted that the NCDS slightly outperforms the CDS due to the fact that the latter requires an additional OFDM symbol for transmitting the reference signals for channel estimation. Then, the proposed fully NCDS system has a significant higher throughput as compared to the fully CDS one, since the NCDS is capable of transmitting more data symbols at the first stage, where no reference signals are needed.

\subsection{Throughput Comparison for different RIS sizes in realistic scenarios}

In Fig. \ref{fig:throughput2}, a throughput comparison for different RIS sizes and training periods is plotted, where the total number of OFDM symbols to be transmitted corresponds to $N=4N_{l}$. Note that the upper-bound is computed using (\ref{eqn:thput_c}) and (\ref{eqn:thput_n}) by setting $P_{e,l}=0$ and $P_{e,h}=0$. Theoretically, a higher number of passive elements of the RIS will provide a more directive beam, and hence, UEs further from the BS will be able to be discovered while closer UEs will face an improvement in their links. However, the simulation results are showing that having a extremely large RIS may even degrade the overall performance in realistic channel environments. 
This can be seen when comparing $M=8\times 8$ and $M=16\times 16$ in this Figure. Certainly, a larger RIS will produce a narrower directive beam enabling a better spatial resolution. However, it requires a higher number of entries for the codebook in order to be able to sweep the whole space. Otherwise, a perfect alignment between the pencil-sharp beam and the strongest cluster of the channel cannot be achieved, and consequently, the resulting reflecting channel gain (BS-RIS-UE) for $M=16\times 16$ is lower than $M=8\times 8$, since the latter is even able to point towards several clusters of the channel, as consequence of having a wider beam. 
Then, when the number of entries in the codebook to be transmitted at the first stage is the same ($N_{CB}=N_{l}$ and $N_{h}=3N_{l}$), the larger RIS shows a slightly worse performance. Additionally, we have to consider that increasing the number of entries in the codebook to be transmitted at the first stage in the figure, the case ($N_{CB}=2N_{l}$ and $N_{h}=2N_{l}$) will enable to explore the whole space with a better spatial resolution, and therefore, the gain of reflective link can be enhanced. Nevertheless, this improvement comes at the expense of reducing the number of OFDM symbols transmitted at the second stage ($N_{h}$), degrading the overall performance of the system.  This explains why in the latter case the performance of the larger RIS is remarkably worse, and highlights the importance of considering the effect of the training on the performance.

\eAddFig{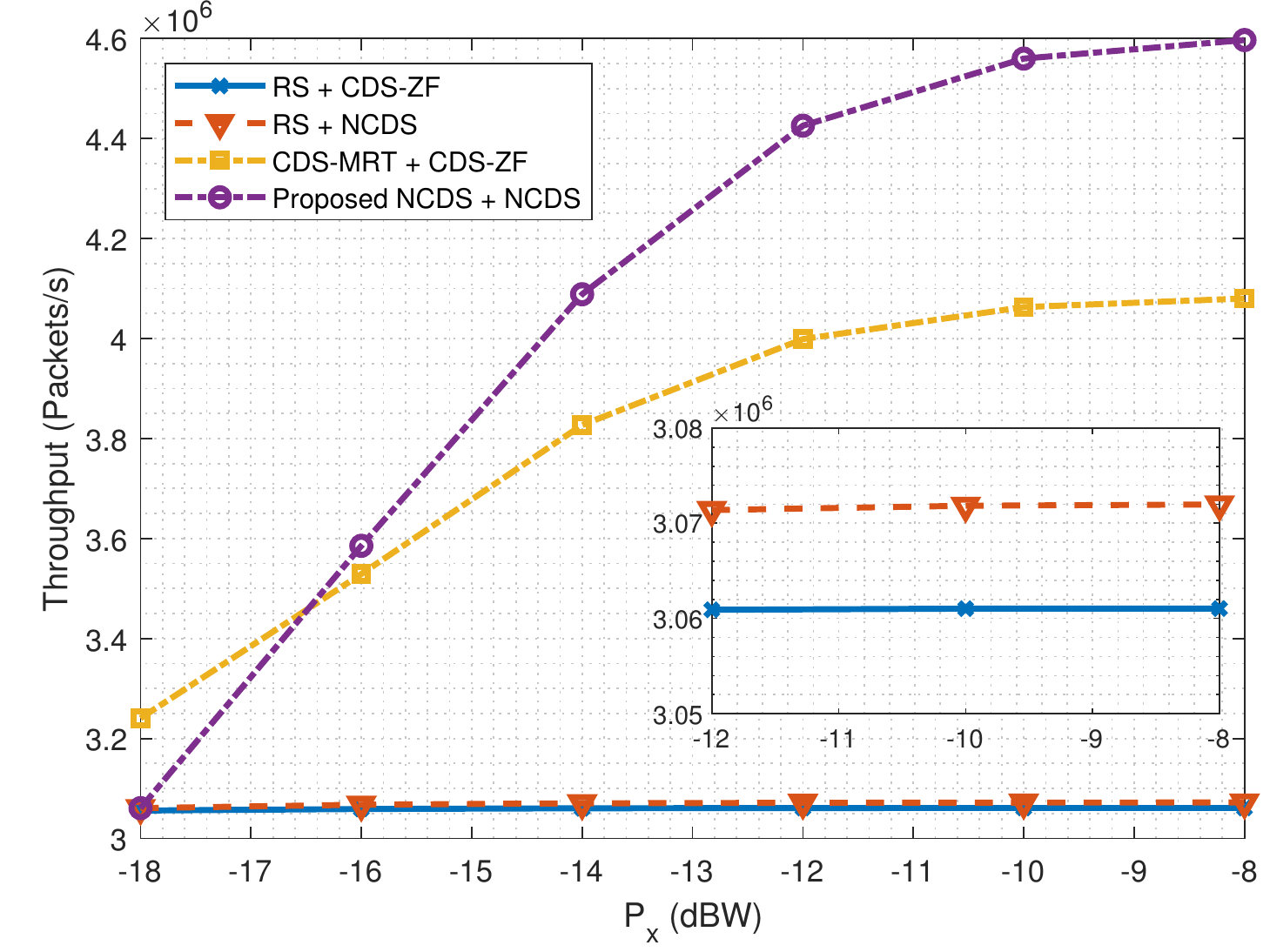}{1}{fig:throughput}{Throughput comparison between CDS and NCDS for a speed of $4$ m/s, which corresponds to $N=2N_{l}$.}
\eAddFig{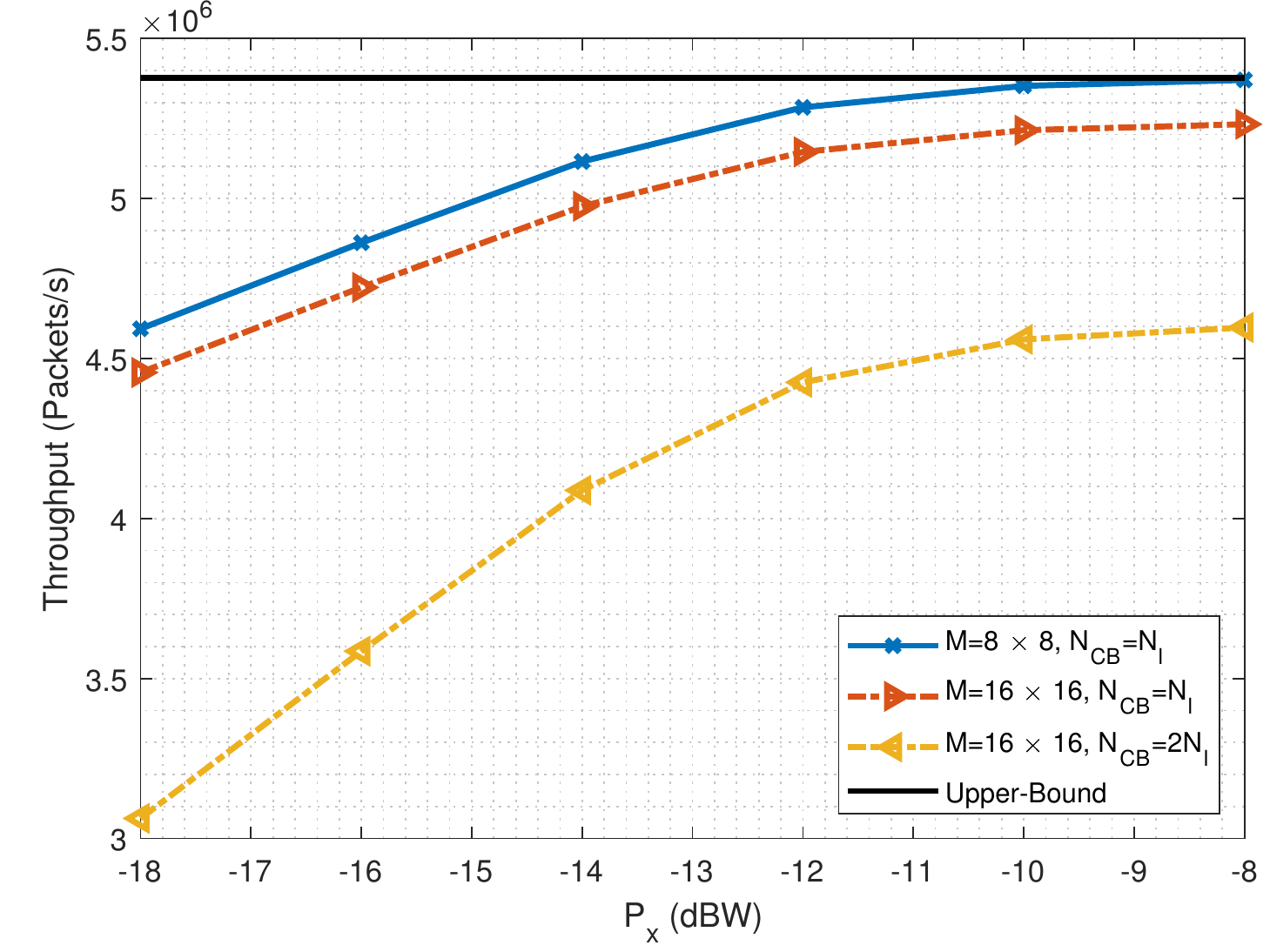}{1}{fig:throughput2}{Throughput comparison for different RIS size ($M=8\times 8$ and $16\times 16$) and training periods ($N_{CB}=N_{l}$ and $N_{CB}=2N_{l}$) and $N=4N_{l}$.}

\section{Conclusions}\label{sec:conclusion}
This paper investigated NCDS based on differential modulation adequately combined with a codebook-based beam training of the RIS (NCDS-RIS) in order to provide a zero overhead training procedure. The proposed scheme is able to execute the beam training process at the RIS, and at the same time, the differential data symbols are transmitted using the direct and reflective links, where the latter is strongly time-varying as a consequence of the beam training. Therefore, this combination is able to efficiently obtain the best phase configuration for the RIS and increase the system data-rate, especially in scenarios with moderate to high mobility and/or highly attenuated direct BS-UE links. 

Moreover, the proposed method is simple, yet effective, as compared to conventional RIS-empowered systems based on CDS. In contrast to CDS, the presented analysis of the efficiency and complexity revealed that, the NCDS requires a substantially smaller amount of reference signals and complex products. Hence, these additional benefits will reduce the cost, energy consumption and latency of RIS-empowered communications.

\appendices

\section{Computation of the Average Interference and Noise Power}\label{appendix:geo}
The derivation of closed-form expression for the interference and noise terms required in (\ref{eqn:sinr}) for the considered geometric wideband channel model is presented. It is assumed that the channel frequency responses of any two contiguous subcarriers of the OFDM signal have a similar value as ($\mathbf{h}_{k,n}=\mathbf{h}_{k-1,n}$, $2 \leq k, \leq K$, $1 \leq n, \leq N$). Note that this assumption is widely accepted in multi-carrier waveforms, since the number of subcarriers is typically designed to be much larger than the number of taps of the channel impulse response.

Taking into account the received signal given in (\ref{eqn:model_chan1}), the first term after performing the differential decoding in (\ref{eqn:diff_dec1}) can be decomposed as
\begin{equation} \label{eqn:term1}
	\begin{split}
		I_{1} & = \eherm{\mathbf{h}_{k,n}}\mathbf{h}_{k,n} s_{k,n} = \eherm{\mathbf{h}_{d,k} + \mathbf{h}_{r,k}} \ecs{\mathbf{h}_{d,k} + \mathbf{h}_{r,k}} s_{k,n}^{2} \\
		& = \ecs{\eabsn{\mathbf{h}_{d,k}}{2} + \eabsn{\mathbf{h}_{r,k}}{2}+2\ereal{\mathbf{h}_{d,k}^{H}\mathbf{h}_{r,k}}} s_{k,n}^{2},
	\end{split}
\end{equation}
\begin{equation*}
	1\leq k \leq K, \quad 1 \leq n \leq N,
\end{equation*}
where  it can be easily shown that the three terms are independent to each other, and hence, the average power of (\ref{eqn:term1}) can be decomposed as
\begin{equation} \label{eqn:term1_exp2}
	\begin{split}
		\eexpabstwo{I_{1}} & = P_{x}^{2}\eexpabsn{\mathbf{h}_{d,k}}{4} + P_{x}^{2} \eexpabsn{\mathbf{h}_{r,k}}{4} \\
		& +P_{x}^{2}\eexpabstwo{2\ereal{\mathbf{h}_{d,k}^{H}\mathbf{h}_{r,k}}}.
	\end{split}
\end{equation}

The first term of (\ref{eqn:term1_exp2}) can be obtained as
\begin{equation} \label{eqn:hd4}
	\begin{split}
		\eexpabsn{\mathbf{h}_{d,k}}{4} & = B^{2}L_{d}^{2}\sum_{c=1}^{C_{d}} \ecs{\evar{\eabsn{\alpha_{d,c}}{2}} + \ecs{\eexpabsn{\alpha_{d,c}}{2}}}^{2} \\ 
		& = 2\beta_{d}^{4}B^{2},
	\end{split}
\end{equation}
where the small-fading coefficient of all taps of the direct BS-UE channel ($\mathbf{h}_{d,k}$) are independent random variables to each other, and hence, it satisfies that
\begin{equation}\label{eqn:indep_d}
	\eexp{\alpha_{d,c_{1}}\alpha_{d,c_{2}}} = 0, \quad c_{1} \neq c_{2}, \quad 1 \leq c_{1}, c_{2} \leq C_{d}.
\end{equation}
The gain produced by the array of the BS after performing the differential modulation is obtained by
\begin{equation}\label{eq:array_gain}
	\mathbf{a}_{\rm BS}^{H}\ecs{\phi_{d,c},\theta_{d,c}}\mathbf{a}_{\rm BS}\ecs{\phi_{d,c},\theta_{d,c}} = B, \quad 1 \leq c \leq C_{d},
\end{equation}
and $\eabsn{\alpha_{d,c}}{2}$ is a random variable that follows
\begin{equation}\label{eq:exp_distri}
	\eabsn{\alpha_{d,c}}{2}\sim\eexpod{\sigma_{d,c}^{-2}}.
\end{equation}

The second term of (\ref{eqn:term1_exp2}) corresponds the reflected BS-UE channel via RIS after choosing the phase configuration ($\mathbf{h}_{r,k}$). Assuming that the RIS is capable of pointing the narrow directive beams towards the LoS components of the BS-RIS and RIS-UE channels, and the NLoS components are spatially filter out, the reflective channel can be considered a deterministic value as
\begin{equation} \label{eqn:hr4}
	\eexpabsn{\mathbf{h}_{r,k}}{4} = \beta_{r}^{4}B^{2},
\end{equation}
while last term of (\ref{eqn:term1_exp2}) can be expressed as
\begin{equation} \label{eqn:hdhr}
	\begin{split}
		& \eexpabsn{2\ereal{\mathbf{h}_{d,k}^{H}\mathbf{h}_{r,k}}}{2} = \evar{2\ereal{\mathbf{h}_{d,k}^{H}\mathbf{h}_{r,k}}} = \\
		&  = 4 \beta_{r}^{2}B\evar{\ereal{\mathbf{h}_{d,k}}} = 2\beta_{d}^{2}\beta_{r}^{2}B^{2}.
	\end{split}
\end{equation}
Making use (\ref{eqn:hd4})-(\ref{eqn:hdhr}), (\ref{eqn:term1_exp2}) can be obtained as
\begin{equation} \label{eqn:term1_squa}
	\eexpabstwo{I_{1}}= \ecs{2\beta_{d}^{4}+\beta_{r}^{4}+4\beta_{d}^{2}\beta_{r}^{2}}B^{2}P_{x}^{2}.
\end{equation}

The second and third terms after performing the differential modulation given in (\ref{eqn:diff_dec1}) and (\ref{eqn:diff_dec3}) can be easily obtained since the transmitted symbol, the channel frequency response and noise are independent random variables. Hence, the average power of $I_{2}$ and $I_{3}$ can be computed as
\begin{equation}\label{eq:exp_scn}
	\begin{split}
		\eexpabsn{I_{2}}{2} & = \eexpabsn{I_{3}}{2} \\
		& = \eexpabsn{\mathbf{h}_{k,n}}{2} \eexpabsn{x_{k,n}}{2} \eexpabsn{\mathbf{v}_{k-1,n}}{2} \\
		& = \sigma_{v}^{2}P_{x}\ecs{\beta_{d}^{2}+\beta_{r}^{2}}.
	\end{split}
\end{equation}

Finally, the fourth term after performing the differential modulation given in (\ref{eqn:diff_dec3}) can be straightforwardly computed since the noise samples at two different subcarriers are independent random variables, therefore
\begin{equation}\label{eq:exp_nn}
	\eexpabsn{I_{4}}{2} = \eexpabsn{\mathbf{v}_{k-1,n}}{2}\eexpabsn{\mathbf{v}_{k,n}}{2} = \sigma_{v}^{4}.
\end{equation}

%

\ifCLASSOPTIONcaptionsoff
\newpage
\fi
\bibliographystyle{IEEEtran}
\bibliography{./bibtex/IEEEabrv,./bibtex/IEEEexample}{}

\begin{IEEEbiography}[{\includegraphics[width=1in,height=1.25in,clip,keepaspectratio]{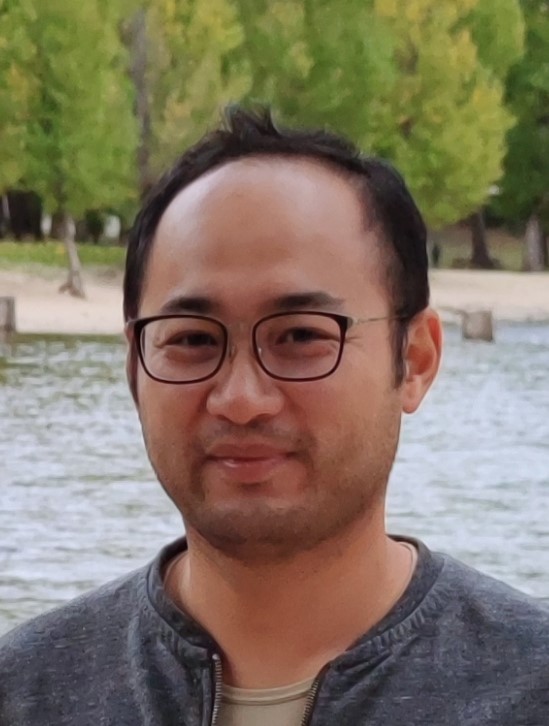}}]{Kun Chen-Hu} (S'16-GS'20-M'21) received his Ph.D. degree in Multimedia and Communications in 2019 from Universidad Carlos III de Madrid (Spain). Currently, he is a post-doctoral researcher in the same institution. He was awarded by UC3M in 2019 recognizing his outstanding professional career after graduation. He visited Eurecom (France) and Vodafone Chair TU Dresden (Germany), both as guest researcher. He also participated in different research projects in collaboration with several top companies in the area of mobile communications. He is the Web Chair for Globecom 2021, Madrid (Spain), and online content editor for IEEE ComSoc. His research interests are related to signal processing techniques, such as waveforms design, reconfigurable intelligent surfaces, non-coherent massive MIMO and channel estimation.
\end{IEEEbiography}

\begin{IEEEbiography}[{\includegraphics[width=1in,height=1.25in,clip,keepaspectratio]{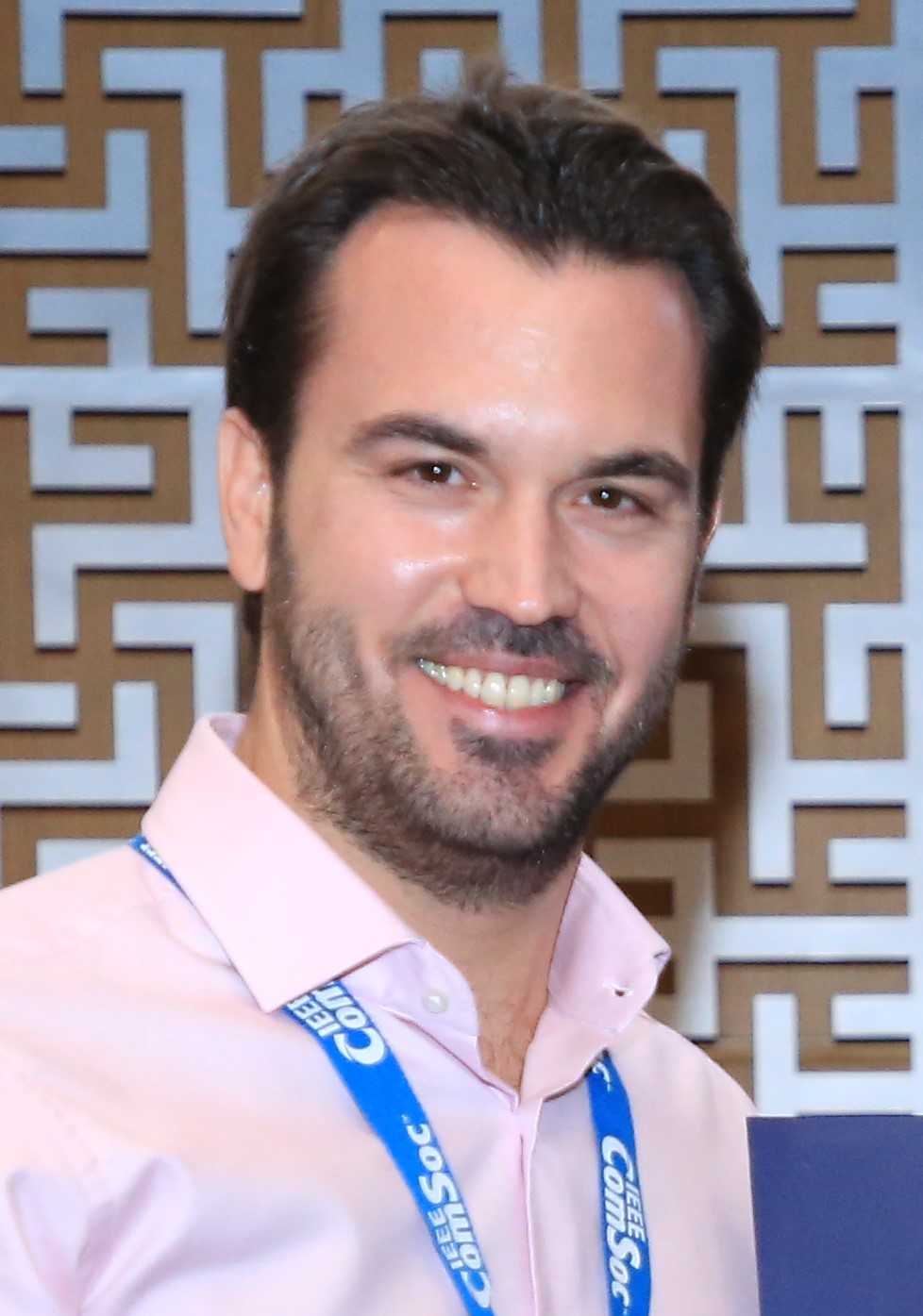}}]{George C. Alexandropoulos} received the Engineering Diploma, M.A.Sc., and Ph.D. degrees in Computer Engineering and Informatics from the the School of Engineering, University of Patras, Greece in 2003, 2005, and 2010, respectively. He has held research positions at various Greek universities and research institutes, as well as at the Mathematical and Algorithmic Sciences Lab, Paris Research Center, Huawei Technologies France, and he is currently an Assistant Professor with the Department of Informatics and Telecommunications, School of Sciences, National and Kapodistrian University of Athens (NKUA), Greece. He also serves as a Principal Researcher at the Technology Innovation Institute, Abu Dhabi, United Arab Emirates. His research interests span the general areas of algorithmic design and performance analysis for wireless networks with emphasis on multi-antenna transceiver hardware architectures, active and passive reconfigurable metasurfaces, integrated communications and sensing, millimeter wave and THz communications, as well as distributed machine learning algorithms. He currently serves as an Editor for IEEE Wireless Communications Letters, ELSEVIER Computer Networks, Frontiers in Communications and Networks, and the ITU Journal on Future and Evolving Technologies. In the past, he has held various fixed-term and guest editorial positions for IEEE Transactions on Wireless Communications and IEEE Communications Letters, as well as for various special issues at IEEE journals. Prof. Alexandropoulos is a Senior Member of the IEEE Communications, Signal Processing, and Information Theory Societies as well as a registered Professional Engineer of the Technical Chamber of Greece. He is also a Distinguished Lecturer of the IEEE Communications Society. He has participated and/or technically managed more than 10 European Union (EU) research and innovation projects, as well as several Greek and international research projects. He is currently NKUA's principal investigator for the EU H2020 RISE-6G research and innovation project dealing with RIS-empowered smart wireless environments. He has received the best Ph.D. thesis award 2010, the IEEE Communications Society Best Young Professional in Industry Award 2018, the EURASIP Best Paper Award of the Journal on Wireless Communications and Networking 2021, the IEEE Marconi Prize Paper Award in Wireless Communications 2021, and a Best Paper Award from the IEEE GLOBECOM 2021. More information is available at \href{www.alexandropoulos.info}{www.alexandropoulos.info}
\end{IEEEbiography}

\begin{IEEEbiography}[{\includegraphics[width=1in,height=1.25in,clip,keepaspectratio]{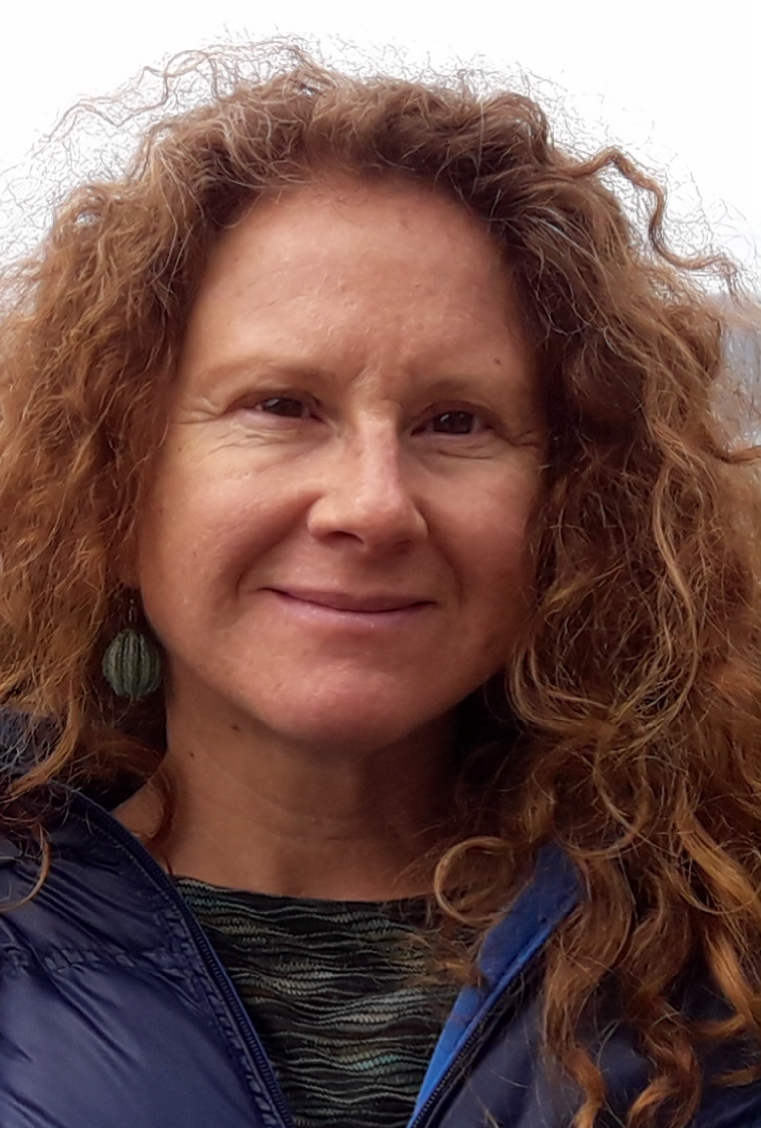}}]{Ana García Armada} (S’96-A’98-M’00-SM’08) is currently a Professor at Universidad Carlos III de Madrid, Spain, where she is leading the Communications Research Group. She has been visiting scholar at Stanford University, Bell Labs and University of Southampton. Her research has resulted in 9 book chapters, and more than 200 publications in international journals and conferences, as well as 5 patents. She is an Editor of IEEE Transactions on Communications, Area Editor of IEEE Open Journal of the Communications Society, Editor of the ITU Journal on Future and Evolving Technologies and is a regular member of the technical program committees of the most relevant international conferences in his field. She has been part of the organizing committee of the IEEE Globecom 2019 and 2021 (General Chair), IEEE Vehicular Technology Conference Spring 2018, 2019 and Fall 2018, IEEE 5G Summit 2017, among others. She has been Secretary and is now Vice-chair of the IEEE ComSoc Signal Processing and Computing for Communications Committee. She has been Secretary and Chair of the IEEE ComSoc Women in Communications Engineering Standing Committee. She has been Member at Large of the Board of Governors and the Director of Online Content of the IEEE Communications Society from 2020-2021. From January 2022 she is the Vice President of Member and Global Activities of this society. She has held various management positions at Universidad Carlos III de Madrid, including: Deputy Director of the Telecommunications Engineering degree, Deputy Vice Chancellor for International Relations and Director of the Department of Signal Theory and Communications, among others. She has received the Award of Excellence from the Social Council and the Award for Best Teaching Practices from Universidad Carlos III de Madrid, as well as the third place Bell Labs Prize 2014, the Outstanding Service Award 2019 from the SPCE committee of the IEEE Communications Society, the Outstanding Service Award 2020 from the Women in Communications Engineering (WICE) standing committee, and the 2022 IEEE ComSoc/KICS Exemplary Global Service Award.

\end{IEEEbiography}

\end{document}

%% file: NewCommands.tex
\newcommand{\eexpo}[1]{\exp \left(  #1 \right) }
\newcommand{\esine}[1]{\sin \left(  #1 \right) }
\newcommand{\ecosi}[1]{\cos \left(  #1 \right) }

\newcommand{\eabsn}[2]{\left|  #1 \right| ^{#2}}
\newcommand{\eherm}[1]{\left(  #1 \right)^{H}}

\newcommand{\einv}[1]{\left( #1 \right)^{-1}}

\newcommand{\eelem}[3]{\left[  #1 \right]^{#2}_{#3} }

\newcommand{\enormn}[3]{\left| \left| #1\right| \right| ^{#2}_{#3}}

\newcommand{\ecsizeo}[1]{\in\mathbb{C}^{#1}}

\newcommand{\ecsize}[2]{\in\mathbb{C}^{#1\times#2}}

\newcommand{\egausd}[2]{\mathcal{CN}\left( #1,#2\right) }
\newcommand{\eexpod}[1]{\text{Exp}\left( #1\right) }

\newcommand{\eexpabstwo}[1]{\mathbb{E} \left\lbrace  \left| #1 \right|^2 \right\rbrace}
\newcommand{\eexpabsn}[2]{\mathbb{E} \left\lbrace  \left| #1 \right|^{#2} \right\rbrace}
\newcommand{\eexpnortwo}[1]{\mathbb{E} \left\lbrace  \left\lVert #1 \right\rVert^2_F \right\rbrace}
\newcommand{\eexp}[1]{\mathbb{E} \left\lbrace #1 \right\rbrace}
\newcommand{\evar}[1]{\text{Var} \left\lbrace #1 \right\rbrace}

\newcommand{\ereal}[1]{\Re \left\lbrace #1 \right\rbrace}
\newcommand{\eimag}[1]{\Im \left\lbrace #1 \right\rbrace}

\newcommand{\ecs}[1]{\left(   #1 \right)  }

\def\mydate{\leavevmode\hbox{\the\year/\twodigits\month/\twodigits\day}}
\def\twodigits#1{\ifnum#1<10 0\fi\the#1}

\newcommand{\eAddFig}[4]{
	\begin{figure}[!t]
		\centering
		\includegraphics[width=#2\linewidth]{#1}
		\caption{#4}
		\label{#3}
	\end{figure}
}

\newcommand{\eAddTwoColFig}[4]{
	\begin{figure*}[!t]
		\centering
		\includegraphics[width=#2\linewidth]{#1}
		\caption{#4}
		\label{#3}
	\end{figure*}
}